\documentclass[fleqn,usenatbib]{mnras}
\usepackage{mathptmx}
\usepackage[T1]{fontenc}
\usepackage{ae,aecompl}
\usepackage{natbib}
\usepackage{amsmath}
\usepackage{amssymb}	
\usepackage{graphicx}
\usepackage[dvipsnames]{xcolor}
\usepackage{units}
\usepackage{ulem}
\usepackage{fixltx2e}
\DeclareGraphicsExtensions{.eps,.pdf,.png,.jpg}

\usepackage{deluxetable}

\usepackage{color}

\definecolor{darkgreen}{rgb}{0.13, 0.55, 0.13}

\def\hii{H~\textsc{ii}}
\def\harm{\textit{HARM${^2}$}}
\def\orion{\texttt{ORION}}
\def\flash{\texttt{FLASH}}
\def\nort{\texttt{LamFLD}}
\def\noturb{\texttt{LamRT+FLD}}
\def\turb{\texttt{TurbRT+FLD}}
\def\noturblr{\texttt{LamRT+FLD\_LR}}

\def\lesssim{\mathrel{\hbox{\rlap{\hbox{\lower3pt\hbox{$\sim$}}}\hbox{\raise2pt\hbox{$<$}}}}}
\def\gtrsim{\mathrel{\hbox{\rlap{\hbox{\lower3pt\hbox{$\sim$}}}\hbox{\raise2pt\hbox{$>$}}}}}

\title[How Massive Stars get their Mass]{An Unstable Truth: How Massive Stars get their Mass}

\author[A.L. Rosen et al.]{Anna L. Rosen,$^1$\textsuperscript{\thanks{E-mail: alrosen@ucsc.edu}} Mark R. Krumholz$,^{2}$ Christopher F. McKee$,^{3,4}$ Richard I. Klein$^{3,5}$ \\
$^1$Department of Astronomy \& Astrophysics, University of California, Santa Cruz, CA 95064 USA \\
$^2$Research School of Astronomy \& Astrophysics, Australian National University, Canberra, ACT 2611, Australia \\
$^3$Department of Astronomy, University of California, Berkeley, CA 94720, USA \\
$^4$Physics Department, University of California, Berkeley, CA 94720, USA \\
$^5$Lawrence Livermore National Laboratory, PO Box L-23, Livermore, CA 94550, USA\\
}

\date{Submitted 2016 July 07}
\pubyear{2016}

\begin{document}
\label{firstpage}
\pagerange{\pageref{firstpage}--\pageref{lastpage}}
\maketitle
\begin{abstract}
The pressure exerted by massive stars' radiation fields is an important mechanism regulating their formation. Detailed simulation of massive star formation therefore requires an accurate treatment of radiation. However, all published simulations have either used a diffusion approximation of limited validity; have only been able to simulate a single star fixed in space, thereby suppressing potentially-important instabilities; or did not provide adequate resolution at locations where instabilities may develop. To remedy this we have developed a new, highly accurate radiation algorithm that properly treats the absorption of the direct radiation field from stars and the re-emission and processing by interstellar dust. We use our new tool to perform three-dimensional radiation-hydrodynamic simulations of the collapse of massive pre-stellar cores with laminar and turbulent initial conditions and properly resolve regions where we expect instabilities to grow. We find that mass is channeled to the stellar system via gravitational and Rayleigh-Taylor (RT) instabilities, in agreement with previous results using stars capable of moving, but in disagreement with methods where the star is held fixed or with simulations that do not adequately resolve the development of RT instabilities. For laminar initial conditions, proper treatment of the direct radiation field produces later onset of instability, but does not suppress it entirely provided the edges of radiation-dominated bubbles are adequately resolved. Instabilities arise immediately for turbulent pre-stellar cores because the initial turbulence seeds the instabilities. Our results suggest that RT features are significant and should be present around accreting massive stars throughout their formation.\\
\end{abstract}


\section{Introduction}
\label{sec:intro}
Massive stars live fast and die young. They are the major contributors to heavy element production in the Universe through their explosive deaths enriching the interstellar medium (ISM). Massive stars are rare, representing only $\sim 1\%$ of the stellar population by number, yet they dominate the energy budget in the Milky Way and other star-forming galaxies because of their strong radiation fields, stellar winds, and supernova explosions. This stellar feedback -- the injection of energy and momentum by stars into the ISM -- limits their masses thereby affecting nuclear yields, slows down nearby star formation, and affects galaxy evolution. 

Recent studies suggest that the pressure exerted by massive stars' radiation fields may be the dominant feedback mechanism during their formation \citep{Krumholz2009a, Kuiper2011a, Kuiper2012a, Klassen2016a}. Massive stars have short Kelvin-Helmholtz timescales (the time required for a star to radiate away its gravitational binding energy) and contract to the main-sequence while they are accreting \citep{Palla91a, Palla92a, Behrend01a, Hosokawa2009a}. Therefore they attain their main sequence luminosities while they are still actively accreting and the radiation pressure associated with their high luminosities can oppose gravity and halt accretion \citep{Larson71a, Yorke79a, Yorke95b, Wolfire86a, Wolfire87a, Yorke1999a}. 

The relative importance of the radiative force ($f_{\rm rad}$) and the gravitational force ($f_{\rm grav}$) can be described in terms of the Eddington ratio, $f_{\rm edd} = f_{\rm rad}/f_{\rm grav}$, which simplifies to
 \begin{equation}
\label{eqn:feddh}
f_{\rm edd} = 7.7 \times 10^{-5} \left( 1 + f_{\rm trap} \right)  \left( \frac{L_{\rm \star}}{M_{\rm \star}} \right)_{\rm \odot} \left( \frac{\Sigma}{1 \; \rm{g \; cm^{-2}}} \right)^{-1}
\end{equation}
where $\Sigma$ is the surface density of the optically thick infalling material and $\left(L_{\rm \star}/M_{\rm \star}\right)_{\rm \odot}$ is the stellar light-to-mass ratio in solar units. The factor $\left( 1 + f_{\rm trap} \right)$ included in $f_{\rm rad}$ denotes the combined contribution from the direct radiation pressure associated with the first absorption of the stellar radiation field and the reprocessed thermal, diffuse radiation pressure associated with the re-emission by interstellar dust, respectively. Here $f_{\rm trap}$ denotes the trapping factor at which the radiation field is enhanced by the subsequent absorption and re-emission by interstellar dust. For spherically symmetric accretion, Equation (\ref{eqn:feddh}) exceeds unity for stars with masses above $\sim 15-20 \;  M_{\rm \odot}$ \citep{Pollack1994a,Krumholz2009a}. If accretion onto the star were isotropic then stars with masses in excess of this limit should not form, a  problem commonly known as ``the radiation pressure barrier problem."  However, recent studies suggest that massive stars with initial masses well in excess of $150 \; M_{\rm \odot}$ exist and can have a dramatic impact on their environments \citep{Crowther2010a, Crowther2016a}. 

Given the existence of massive stars, a number of solutions to the radiation pressure problem have been proposed in the literature. \citet{Nakano89a} and \citet{Jijina96a} present analytic models suggesting that accretion through a disk could circumvent the radiation pressure barrier, while \citet{McKee2003a} suggest that high accretion rates could provide sufficient ram pressure even in spherical symmetry. \citet{Krumholz2005c} showed that escape of radiation through outflow channels could ease the radiation pressure problem. Numerical simulations within the last several decades generally support these hypotheses. Most of these simulations model the collapse of isolated, slowly rotating, and initially laminar pre-stellar massive cores \citep{Yorke1999a, Yorke2002a, Krumholz2009a, Kuiper2011a, Kuiper2012a, Klassen2016a}. In these idealized simulations, the radiation pressure barrier is circumvented by the formation of an optically thick accretion disk that surrounds the massive star. With this anisotropy, the radiative flux easily escapes along the polar directions of the star, launching radiation pressure dominated bubbles both above and below the star. This ``flashlight" effect allows material to be funneled to the star by the accretion disk and gravitational instabilities present in the disk can enhance the accretion rate onto the star \citep{Yorke2002a, Krumholz2009a, Kuiper2011a, Kuiper2012a, Klassen2016a}.

Whether material is supplied to the star via disk accretion alone has been heavily debated in the literature \citep{Krumholz2009a, Kuiper2011a, Kuiper2012a, Klassen2016a}. \citet{Krumholz2009a} performed the first adaptive mesh refinement (AMR) 3D radiation-hydrodynamic simulation of the formation of a massive stellar system and found that the dense shells that surround the radiation pressure dominated bubbles become radiative Rayleigh-Taylor (RT) unstable. In this configuration, the dense shells that surround the rarefied radiation pressure dominated bubbles develop perturbations at the interface that grow exponentially, leading to ``fingers" in the heavier fluid (the accreting gas) that sink into the lighter, more buoyant fluid (represented by the radiation field; \citet{Jacquet2011a}). These RT ``fingers" can reach the star-disk system if they are not pushed back by radiation pressure, and deliver a significant amount of mass to the accretion disk that can then be incorporated into the star. 

The presence of these instabilities can allow stars to grow beyond their Eddington limit but their development and growth is sensitive to how the radiation pressure is treated. \citet{Krumholz2009a} only included the dust-reprocessed radiation pressure, which was modeled with the gray flux limited diffusion (FLD) approximation, and assumed that the stellar radiation energy was depositied within the vicinity of the star, which underestimated the true radiation pressure. If the radiation pressure, especially the component of the radiative force that is anti-parallel to the gravitational force, is underestimated then the gas is less likely to be pushed away by radiation. Furthermore, an anisotropic radiation field can lead to density perturbations in the dense shells of the radiation pressure dominated bubbles that can then amplify and become RT unstable. These instabilities can grow and deliver material to the star-disk system.

To better represent the true radiation field in massive star formation simulations \citet{Kuiper2010a} developed a hybrid radiation algorithm that included a multi-frequency raytracer, in which a series of rays travel radially away from the star and transfer energy and momentum to the absorbing dust, coupled to gray FLD to model the diffuse dust-reprocessed radiation field. With this method, \citet{Kuiper2011a, Kuiper2012a} performed a series of 3D simulations of the formation of massive stars from the collapse of laminar pre-stellar cores on a non-adaptive spherical non-uniform grid with resolution increasing logarithmically towards the center. The authors find that the star is fed through disk accretion only and that the radiation pressure dominated bubbles do not become RT unstable. They conclude that inclusion of the direct radiation pressure is responsible for maintaining stability of the expanding bubble shells.

The work of \citet{Krumholz2009a} and \citet{Kuiper2011a, Kuiper2012a} both have their advantages and disadvantages. AMR simulations with a general Cartesian geometry, such as the simulation presented in \citet{Krumholz2009a}, can handle an arbitrary number of moving stars. The resulting gravitational interaction of  the massive star with its accretion disk can induce gravitational instabilities leading to disk fragmentation. In addition, movement of the massive star within the accretion disk can lead to shielding of the stellar radiation field resulting in a greater asymmetry in the direct radiation pressure, potentially seeding RT instabilities. One key advantage in AMR simulations, as compared to a non-adaptive grid, is that instabilities that may develop in the dense bubble shells can be resolved dynamically throughout the bubble evolution. In classical RT theory, the smallest perturbations grow fastest in the linear regime and these perturbations can only grow if they are resolved. The bubble shells in the work of \citet{Krumholz2009a} are resolved to the finest level, likely allowing for small RT instabilities to grow large enough to deliver material to the star-disk system. 

In contrast, the bubble shells in the work of \citet{Kuiper2011a, Kuiper2012a} are poorly resolved because they use a non-adaptive spherical grid. Furthermore, the star is artificially held at the origin of the grid, thereby suppressing potentially-important instabilities that could seed RT instabilities. However, these simulations included a much better treatment of the radiation field by incorporating a multi-frequency raytracer to model the direct radiation field. In such a geometry raytracing becomes trivial because the rays travel radially from the non-moving star, but this geometry can not support additional stars or disk asymmetries induced by stellar movement. Hence, the next generation of massive star formation simulations must include the advantages of both methods to better understand how massive stars can overcome the Eddington limit by including hybrid radiative transfer on adaptive grids.

The question of whether RT instability is important for massive star formation has been muddied further by studies of radiation pressure-driven instabilities in the context of galactic winds. \citet{Krumholz2012b, Krumholz2013a} study the ability of radiation to drive galactic winds using the same FLD methods as \citet{Krumholz2009a}, and find that RT instabilities arise and prevent the onset of winds entirely. \citet{Rosdahl15a} reach the same conclusion using an M1 closure to treat the radiation. \citet{Davis2014a}, using a variable Eddington tensor method on a fixed grid, and \citet{Tsang15a}, using implicit Monte Carlo, concur that RT instability occurs, but find that it does not prevent a wind from being launched, contrary to the results of \citeauthor{Krumholz2012b} and \citeauthor{Rosdahl15a}. Moreover, none of these calculations included a treatment of the direct radiation field.

The conflicting results discussed thus far have motivated the implementation of a new generation of hybrid radiation solvers in AMR simulation codes. Both \citet{Klassen2014a} and \citet{Rosen2016a} developed novel hybrid radiation schemes in the \flash\  and \orion\ AMR simulation codes, respectively. Both implementations model the direct radiation field with a raytracer while the diffuse component is handled by a FLD solver, and can be used with an arbitrary number of moving stars. The raytracer employed in the Hybrid Adaptive Ray-Moment Method (\harm) algorithm developed by \citet{Rosen2016a} uses the method of long characteristics, which traces rays on a cell by cell basis thus providing maximum possible accuracy. Their method is adaptive, in which rays are allowed to split as they travel away from their source, greatly reducing the computational cost; and is capable of representing multi-frequency stellar irradiation \citep{Abel2002a, Wise2011a, Rosen2016a}. The multi-frequency treatment  is ideal for stars since they have color temperatures much higher than the absorbing medium. The raytracer employed in \citet{Klassen2014a} models only single frequency irradiation and uses hybrid characteristics, which is a combination of long characteristics within individual grids and short characteristics between grids (i.e., in which only neighboring grid cells are used to interpolate incoming intensities; \citet{Rijkhorst2006a}). The method of short characteristics is typically faster but more diffusive than long characteristics. Because of this limitation the long characteristics method employed in \citet{Rosen2016a} has been highly optimized.

To revisit the problem of massive star formation and whether or not mass is delivered to the star via RT instabilities, \citet{Klassen2016a} simulated the collapse of initially laminar pre-stellar cores with the new hybrid radiation algorithm presented in \citet{Klassen2014a}. Like the work of \citet{Kuiper2011a, Kuiper2012a} they find that their radiation pressure dominated bubbles remain stable and that the massive star is fed by disk accretion alone. However, the authors employ poor refinement criteria in their simulations, which results in the bubble shells being poorly resolved, potentially suppressing RT instabilities that are not resolved. To address this, we perform similar simulations of the collapse of a laminar massive pre-stellar core in which we choose to resolve the bubble shells, like that of \citet{Krumholz2009a}, and use the \harm\ hybrid radiation algorithm to determine if RT instabilities are a real effect or if the direct radiation pressure inhibits their growth. As we will show, the development of RT instabilities is resolution dependent and therefore we find that authors can arrive at conflicting results if the bubble shells are not properly refined.

The simulations discussed thus far were highly idealized. To date only the collapse of initially laminar massive pre-stellar cores have been studied numerically with a detailed treatment of the direct and diffuse radiation fields, yet observations of star forming regions show that star-forming cores are turbulent \citep{Tatematsu2008a,Sanchez-Monge2013a}. In such a configuration, the initial turbulence should act as seeds for RT instabilities. Furthermore, the asymmetric gas distribution in turbulent cores can yield low-density channels where radiation can easily escape, even in the absence of channels cut by outflows.

The purpose of this paper is to study how radiation pressure affects the formation of massive stars via direct numerical simulation. For this work, we use the new highly accurate \harm\ algorithm described in \citet{Rosen2016a}, which treats the direct radiation field from stars and the indirect radiation field associated with the re-emission and processing by interstellar dust.  In this work, we simulate the collapse of both initially laminar and turbulent pre-stellar cores to determine how massive stars attain their mass. For the laminar cores, we also examine how resolution and treatment of radiation pressure can affect the onset of RT instabilities. We simulate the collapse of an initially turbulent core to model a more realistic setup of how massive stars form to show that RT instabilities are a common occurrence in their formation. The simulations presented in this work are still highly idealized since we do not include magnetic fields or outflows. This paper is organized as follows: we describe our numerical methodology and simulation design in Section \ref{sec:numeth}, we present and discuss our results in Sections \ref{sec:results} and \ref{sec:discussion}, respectively, and conclude in Section \ref{sec:conclusion}.

\begin{figure}
\centerline{\includegraphics[trim=0.2cm 0.2cm 0.2cm 0.2cm,clip,width=0.95\columnwidth]{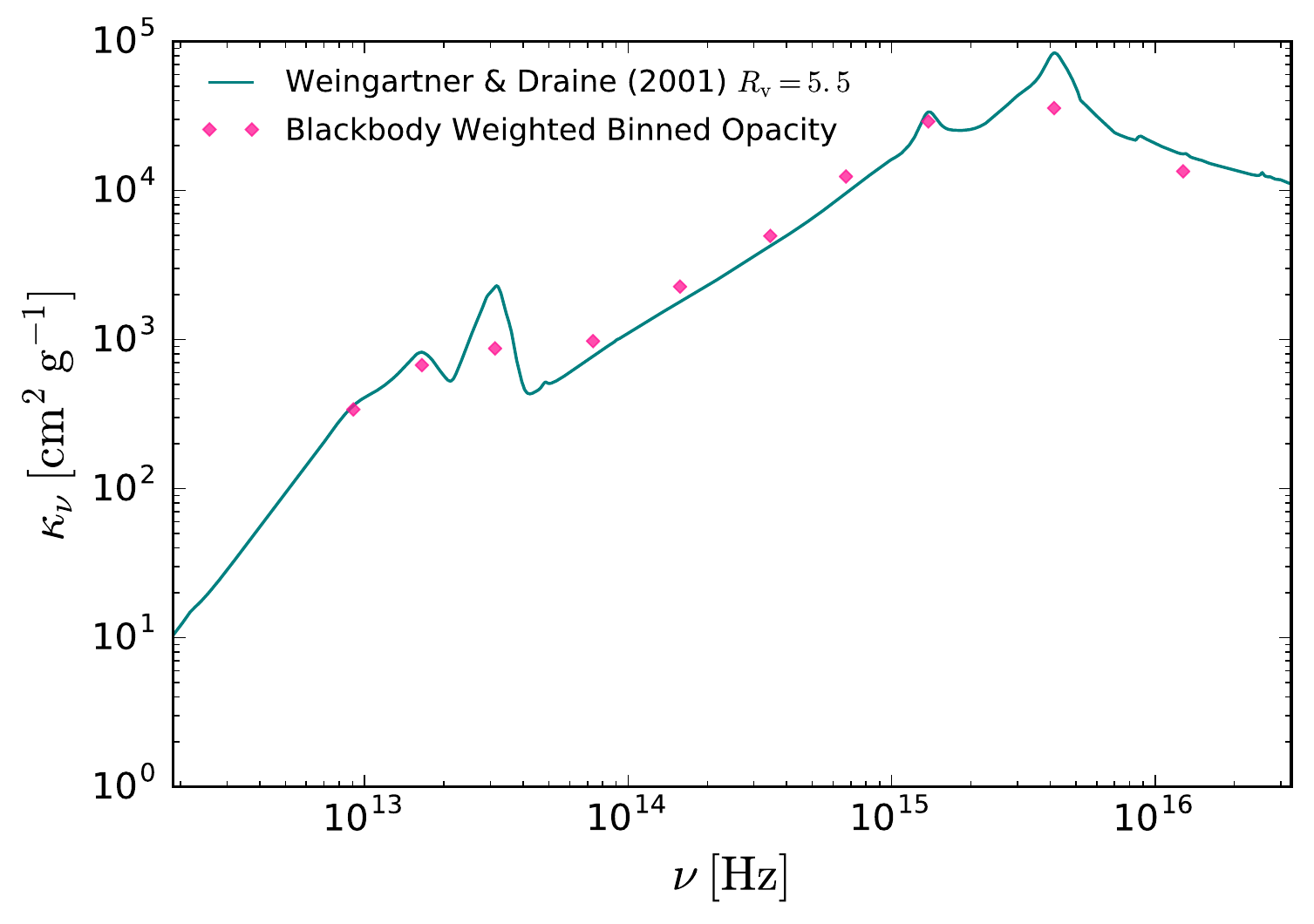}}
\centering
\caption{
Specific frequency dependent dust opacities (per gram of dust) from \citet{Weingartner2001a} for their $R_{\rm v}=5.5$ extinction curve (teal line) with black body weighted binned opacities (pink diamonds) over-plotted for ten frequency bins used in the simulations presented in this work.
}
\label{fig:opacity}
\end{figure}

\section{Numerical Method}
\label{sec:numeth}
In this paper, we simulate the collapse of isolated laminar and turbulent massive pre-stellar cores with the \orion\ adaptive mesh refinement (AMR) code. \orion\ includes  \citep{klein1999a},  radiative transfer \citep{Howell2003a, Krumholz2007a, Shestakov2008a, Rosen2016a}, self-gravity \citep{truelove1998a}, accreting sink particles \citep{Truelove1997a, Krumholz2004a}, a protostellar evolution model used to represent the sink particles as radiating protostars \citep{Offner2009a}, protostellar outflows \citep{Cunningham2011a}, and magnetic fields \citep{Li2012a}. In order to treat both the direct (stellar) and indirect (dust-reprocessed) radiation fields we use the multi-frequency Hybrid Adaptive Ray-Moment Method (\harm) described in \citet{Rosen2016a}, which combines direct solution of the frequency-dependent radiative transfer equation along long characteristics launched from stars to treat the direct stellar radiation field with a gray flux-limited diffusion (FLD) method to treat the radiation field produced by thermal emission from dust \citep{Krumholz2007a}. We describe the equations solved by our code in Section \ref{sec:numerics}, our stellar radiation feedback prescription in Section \ref{sec:starrad}, the initial and boundary conditions for our simulations in Section \ref{sec:ics}, and our refinement criteria and sink creation requirements in Section \ref{sec:sink}.
\subsection{Evolution Equations}
\label{sec:numerics}
\orion\ uses a Cartesian adaptive grid in which every cell has a state vector of conserved quantities ($\rho$, $\rho \mathbf{v}$, $\rho e$, $E_{\rm R}$). Here $\rho$ is the density, $\rho \mathbf{v}$  is the momentum density, $\rho e$ is the total internal plus kinetic gas energy density, and $E_{\rm R}$ is the radiation energy density in the rest frame of the computational domain. In addition to the fluid, \orion\ contains Lagrangian radiating sink particles that accrete from the gas and interact with it via gravity and radiation. The star particles, indexed by subscript $i$, are characterized by their position $\mathbf{x}_{i}$, momentum $\mathbf{p}_i$, mass $M_i$, and luminosity $L_{i}$, as determined by the protostellar evolution model described in \citet{McKee2003a} and \citet{Offner2009a}. They accrete mass, momentum, and energy from the computational grid at rates $\dot{M}_i$, $\dot{\mathbf{p}}_i$, and $\dot{\varepsilon}_i$; the distribution of these quantities over cells in the computational grid is described by a weighting kernel $W(\mathbf{x}-\mathbf{x}_i)$, which is non-zero only within 4 computational zones of each particle. Both the value of $\dot{M}_i$ and $\dot{\mathbf{p}}_i$ and the weighting kernel function are determined via the sink particle algorithm of \citet{Krumholz2004a}. Each star particle also produces a direct radiation field that injects energy and momentum into the gas at a rate per unit volume $\dot{\mathbf{p}}_{{\rm rad},i}$ and $\dot{\varepsilon}_{{\rm rad},i}$; we defer discussion of how these two quantities are computed to Section \ref{sec:starrad}. With these quantities the equations governing the evolution of the RHD fluid-particle system are
\begin{eqnarray}
\label{eqn:com}
\frac{\partial \rho}{\partial t} & = & -\mathbf{\nabla} \cdot \left( \rho \mathbf{v} \right) - \sum_{i}  \dot{M}_i W(\mathbf{x} - \mathbf{x}_i) \\
\frac{\partial \left(\rho \mathbf{v} \right)}{\partial t}  &=& -\mathbf{\nabla} \cdot \left( \rho \mathbf{v}  \bf{v} \right) - \mathbf{\nabla} P - \rho \mathbf{\nabla} \phi - \lambda \mathbf{\nabla} E_{\rm R} \nonumber \\
	&& {} + \sum_i \left[\dot{\mathbf{p}}_{\rm rad, \it i}-\dot{\mathbf{p}}_{\it i}W(\mathbf{x} - \mathbf{x}_i)\right]
\label{eqn:cop}
\\
\frac{\partial \left( \rho e\right)}{\partial t} &=& - \mathbf{\nabla} \cdot \left[ (\rho e + P)\textbf{v} \right] - \rho \mathbf{v} \cdot \mathbf{\nabla} \phi - \kappa_{\rm 0P} \rho(4\pi B - cE_{\rm R} ) \nonumber \\ 
	&&+ \lambda \left( 2 \frac{\kappa_{\rm 0P}}{\kappa_{\rm 0R}} - 1 \right) \mathbf{v} \cdot \mathbf{\nabla} E_{\rm R} - \left(\frac{\rho}{m_{\rm p}} \right)^2 \Lambda(T_{\rm g}) \nonumber \\
	&&+ \sum_i \left[\dot{\mathbf{\varepsilon}}_{\rm rad, \it i}-\dot{\varepsilon}_i W(\mathbf{x} - \mathbf{x}_i) \right]
\label{eqn:coe}
\\
\frac{\partial E_{\rm R}}{\partial t} &=& \mathbf{\nabla} \cdot \left( \frac{c \lambda}{\kappa_{\rm 0R} \rho} \mathbf{\nabla} E_{\rm R} \right) + \kappa_{\rm 0P} \rho \left(4\pi B - c E_{\rm R} \right) \nonumber \\
	&&- \lambda \left( 2 \frac{\kappa_{0 \rm P}}{\kappa_{0 R}} - 1\right) \mathbf{v} \cdot \nabla E_{\rm R} 
	- \nabla \cdot \left( \frac{3 - R_2}{2} \mathbf{v} E_{\rm R}\right) \nonumber \\ 
	& & +  \left(\frac{\rho}{m_{\rm p}} \right)^2 \Lambda(T_{\rm g}) 
\label{eqn:coer}
\end{eqnarray}
Equations (\ref{eqn:com})-(\ref{eqn:coer}) describe conservation of gas mass, gas momentum, gas total energy, and radiation total energy. They include terms describing the exchange of these quantities with the star particles, and exchange of energy between radiation and gas. The gas-radiation exchange terms are written in a mixed-frame formulation that allows conservation of total energy to machine precision \citep{mihalas1982a, Krumholz2007a}. We assume an ideal equation of state so that the gas pressure is 
\begin{equation}
P=\frac{\rho k_{\rm B}T}{\mu m_{\rm H}} = \left( \gamma-1\right) \rho e_{\rm T},
\end{equation}
where $T$ is the gas temperature, $\mu$ is the mean molecular weight,  $\gamma$ is the ratio of specific heats, and $e_{\rm T}$ is the thermal energy of the gas per unit mass. We take $\mu=2.33$  and $\gamma=5/3$ that is appropriate for molecular gas of solar composition at temperatures too low to excite the rotational levels of H$_2$; in practice the exact value of $\gamma$ matters little for our computation, because the gas temperature is set almost entirely by radiative effects, with minimal influence from adiabatic compression or expansion. The fluid is a mixture of gas and dust, and at the high densities that we are concerned with the dust will be thermally coupled to the gas, allowing us to assume that the dust temperature is the same as the gas temperature.

In addition to updating fluid quantities, at each time step we also update the properties of the star particles. These change according to
\begin{eqnarray}
\label{eqn:mdot}
\frac{d M_{\rm i}}{dt} = \dot{M} \\
\label{eqn:dvi}
\frac{d \bf{x}_{i}}{dt} =\frac{\bf{p}_{\rm i}}{M_{\rm i}} \\
\label{eqn:dpi}
\frac{d\bf{p}_i}{dt} = -M_i \nabla \phi + \dot{\bf{p}}_i,
\end{eqnarray}
\noindent
where $\phi$ is the gravitational potential that obeys the Poisson equation including contributions from both the fluid and star particles:

\begin{equation}
\label{eqn:pois}
\nabla^2 \phi = 4 \pi G \left[ \rho + \sum_i M_i \delta(\bf{x} - \bf{x}_i)\right].
\end{equation}

Our sink particle algorithm destroys information within four fine-level cells around each star particle (i.e., the particle's accretion radius) and thus we are unable to properly determine if two sink particles will merge when they approach within one accretion radius of one another (i.e., 80 AU). In light of this limitation, we employ the following merging criteria: when two star particles pass within one accretion radius of each other we merge them together if the smaller particle has a mass less than $0.05 \; M_{\rm \odot}$ \citep{Myers2013a}. This threshold corresponds to the largest plausible mass at which second collapse occurs for the protostar \citep{Masunaga1998a, Masunaga2000a}. At masses lower than this value the protostar represents a hydrostatic core that is several AU in size and will likely be accreted by the more massive star. Larger mass protostars will have collapsed down to sizes of roughly several $R_{\rm \odot}$ and will unlikely merge with the nearby protostar.

Finally, the radiation-specific quantities are the blackbody function $B=c a_{\rm R} T^4/(4\pi)$, the co-moving frame specific Planck- and Rosseland-mean opacities $\kappa_{\rm 0P}$ and  $\kappa_{\rm 0R}$, a dimensionless number $\lambda$ called the flux-limiter, and the Eddington factor $R_2$. The last two quantities appear in Equation (\ref{eqn:coer}) and originate from the FLD approximation, which assumes that the radiative flux in the co-moving frame is related to the gradient of the radiation energy density (Fick's Law)
\begin{equation}
\mathbf{F} = - \frac{c \lambda}{\kappa_{\rm 0R}} \nabla E_{\rm R}.
\end{equation}
\orion\ adopts the \citet{Levermore1981a} approximation for $\lambda$ and $R_2$ as given by
\begin{eqnarray}
\lambda = \frac{1}{R} \left( \coth{R} - \frac{1}{R} \right) \\
R = \frac{|\nabla E_{\rm r}|}{\kappa_{0R} \rho E_{\rm R}} \\
R_2 = \lambda + \lambda^2 R^2.
\end{eqnarray} 
The flux limiter, $\lambda$, has the advantage that in an optically thick medium $\lambda \rightarrow 1/3$, thereby giving $\mathbf{F} \rightarrow -\left[\left(c/3\kappa_{\rm 0R}\right) \nabla E_{\rm r}\right]$, the correct value for diffusion. In an optically thin medium $\lambda \rightarrow \left( \kappa_{\rm 0R} E_{\rm R}/|\nabla E_{\rm R}|\right) \mathbf{n_{\rm R}}$, where $\mathbf{n}_{\rm R}$ is a unit-vector that is anti-parallel to $\nabla E_{\rm R}$, yielding $\mathbf{F} \rightarrow c E_{\rm R} \mathbf{n}_{\rm R}$ for the free-streaming limit \citep{Krumholz2007a}.

\subsection{Treatment of Stellar Radiation}
\label{sec:starrad}
In star-forming environments radiation from stars will be absorbed by the dusty gas and deposit momentum and energy (e.g., see Equations (\ref{eqn:cop})-(\ref{eqn:coer})). The dust, which is highly coupled to the gas, will re-emit thermal radiation at infrared wavelengths and transfer energy and momentum to the gas via collisions. At the high densities with which we are concerned, thermal coupling is strong enough that we can safely assume that the gas and dust are at the same temperature. In order to properly model this, we must know the magnitude and direction of the intervening stellar radiation field. With this in mind we use the new \harm\ algorithm described in \citet{Rosen2016a} to treat the first absorption of the (direct) stellar radiation field from stars and subsequent re-emission of radiation from the fluid.  \harm\ is a new hybrid radiative transfer tool developed for adaptive grids that employs an adaptive long-characteristics ray tracing method, first introduced by \citet{Abel2002a} and extended to adaptive grids by \citet{Wise2011a}, to model the radiative flux from point sources.  It is coupled to a moment method, in our case FLD (e.g., see Section \ref{sec:numerics}), which models the re-processed diffuse radiation field intrinsic to the fluid. In short, \harm\ is used to model both the direct and indirect radiation pressure in numerical simulations. 

The method of long characteristics solves the radiative transfer equation along specific rays on a cell by cell basis that originate from the point source. This method provides the best possible accuracy for the radiative flux for point sources that represent stars because it is less diffusive than short and hybrid characteristic methods \citep{Rijkhorst2006a, Klassen2014a}. \harm\ has the advantage that it can be used to model any number of moving point sources, handles multi-frequency radiation, and is highly parallelizable as compared to previous long-characteristic methods developed for adaptive grids \citep{Wise2011a}. We choose to represent the luminosities of stars by a spectrum of frequency-dependent luminosities rather than a bolometric luminosity, $L_{\rm \star}$, because the color temperatures of stars are much higher than the temperature of the absorbing medium. In what follows, we briefly summarize the basic components of  \harm, and refer the reader to \citet{Rosen2016a} for a full description of the algorithm. 

We describe the deposition of energy and momentum to the fluid from the radiation field of a single star but the generalization to multiple point sources is trivial. Each star has a specific luminosity $L_{\nu,}$ and bolometric luminosity given by $L_{\rm \star}=\int^\infty_0 L_{\rm \nu} d\nu$.  We discretize the stellar spectrum in frequency into $N_{\rm \nu}$ frequency bins, with the $j$th bin covering a range in frequency $(\nu_{j-1/2}, \nu_{j+1/2})$. The luminosity of the point source integrated over the $j$th frequency bin is $L_{\rm \star,j} = \int_{\nu_{j-1/2}}^{\nu_{j+1/2}} L_\nu \, d\nu$ where $\sum L_{\rm \star,j}=L_{\rm \star}$. We choose $N_{\rm \nu}=10$ for the simulations presented in this paper because this number of frequency bins does not significantly increase the cost of the adaptive ray trace (e.g., see Figure 6 of \citet{Rosen2016a}) and provides an adequate frequency sampling of $L_{\rm \nu}$.  The frequency bins were hand-chosen to align with important features of the dust opacity curve as shown in Figure \ref{fig:opacity}.
 
We use the frequency dependent stellar atmosphere profiles from \citet{lejeune1997a} to model the stellar spectrum of stars that form in our simulations. These profiles provide the frequency dependent radiative flux of stars on a grid of values in $\log g$ and $T_{\rm eff}$ space, where $g$ is the star's surface gravity and $T_{\rm eff}$ is the star's surface temperature, both of which are supplied by the sub-grid protostellar model in \orion\ \citep{Offner2009a}. At each raytracing step, we compute $\log g$ and $T_{\rm eff}$ for each star and interpolate between the frequency-dependent stellar atmosphere profiles that match most closely to the star's properties. The accretion of material onto the star will also contribute an accretion luminosity
\begin{equation}
\label{eqn:lacc}
L_{\rm acc} = f_{\rm rad} \frac{G M_{\rm \star}\dot{M}_{\rm \star}}{R_{\rm \star}},
\end{equation}
and we model the accretion luminosity, $L_{\rm acc, \nu}$, as a blackbody with temperature $T_{\rm acc} = \left(L_{\rm acc}/(4\pi R^2_{\rm \star} \sigma) \right)^{1/4}$ such that $L_{\rm acc} = \int^\infty_0 L_{\rm acc, \nu} d\nu$. The resulting luminosity from the star and accretion is  $L_{\rm tot} = \sum^{N_{\rm \nu}}_{j=0} (L_{\rm \star,j }+L_{\rm acc,j})$. The quantity $f_{\rm rad}$ is the fraction of the gravitational potential energy of the accretion flow that is converted to radiation rather than being used to drive a wind or advected into the stellar interior; we adopt $f_{\rm rad}=3/4$, following the standard treatment in \citet{Offner2009a} and this value is reasonably consistent with x-wind models of the launching of stellar outflows \citep{Ostriker1995a}.

We wish to solve the time-independent radiative transfer equation 
\begin{equation}
\label{eqn:rad}
\textbf{n} \nabla I(\nu,\mathbf{n}) = -\kappa(\mathbf{n}, \nu )\rho I(\mathbf{n}, \nu ) + \eta (\mathbf{n}, \nu )\rho
\end{equation}
along specific rays that originate from point sources and transverse the computational domain in the radial direction to model the absorption of the direct radiation field from stars. Here $I(\textbf{n}, \nu)$ is the specific intensity of the stellar radiation field at frequency $\rm \nu$ in direction $\mathbf n$ and $\kappa(\mathbf{n}, \nu )$ and $\eta(\mathbf{n}, \nu )$ are the direction and frequency-dependent specific absorption and emission coefficients. We set $\eta(\mathbf{n}, \nu )$ to zero because the direct radiation field has zero emissivity except at the location of stars. We also neglect the effects of scattering because absorption is the dominant transfer mechanism in these simulations. Finally, we note that we can neglect the time dependence of the radiative transfer equation because the light crossing time of a ray ($t_{\rm lc}$) will be much shorter than the opacity variation time scale (i.e., $t_{\rm lc} \ll \kappa/\left(d\kappa/dt\right))$ for the scales and time steps considered in our simulations.  

We discretize the transfer equation in angle on a series of rays originating at the star and traveling radially outward. Each ray is characterized by a direction $\mathbf{n}$ and solid angle $\Omega_{\rm ray}$ that it subtends. Multiplying both sides of Equation (\ref{eqn:rad}) by $4\pi r^2/\Omega_{\rm ray}$, yields an integrated form of the transfer equation
\begin{equation}
\label{eqn:rtray}
\frac{\partial L_{{\rm ray},j}}{\partial r} = -\kappa_j \rho L_{{\rm ray},j},
\end{equation}
where $L_{{\rm ray},j}(r)$ is the luminosity for the $j$th frequency bin at a distance $r$ from the point source and $\kappa_j$ is the specific absorption opacity for the $j$th frequency bin. This equation is subject to the boundary condition $L_{{\rm ray},j}(0) = L_{\rm tot, \it j}/N_{\rm pix}$, where $N_{\rm pix} = 4\pi/\Omega_{\rm ray}$. In order to reduce cost we initially sample the radiation field for each star with 3072 rays and adaptively split each ray into four sub-rays when the following condition is satisfied
\begin{equation}
\label{eqn:split}
\frac{\Omega_{\rm cell}}{\Omega_{\rm ray}} < \Phi_{\rm c},
\end{equation}
\noindent
where $\Omega_{\rm cell}= (\Delta x/r)^2$ is the solid angle subtended by a cell of linear size $\Delta x$ at a distance $r$ from the point source. The quantity $\Phi_{\rm c}$ is the minimum number of rays required to go through each cell, which we set to $3$ in our simulations. This refinement criterion ensures that the cells that interact with rays are adequately resolved.

Our choice for $\kappa_{\rm j}$ depends on whether the primary absorber is dust or molecular gas. Dust is the primary absorber for gas temperatures below $T_{\rm sub} = 1500$ K (i.e., the temperature at which dust sublimes; \citet{semenov2003a}) while molecular hydrogen is the primary absorber for gas temperatures within $T_{\rm sub} \le T < T_{\rm \hii}$ where $T_{\rm \hii} \approx 10^4$ K is the temperature at which we expect hydrogen to become fully ionized, and thus to have the usual Thompson opacity for electron scattering. If the primary absorber is dust we use the frequency dependent dust opacities from \citet{Weingartner2001a} (their $R_{\rm v}=5.5$ extinction curve)
, e.g., see Figure \ref{fig:opacity}) and assume a constant dust-to-gas ratio of $M_{\rm dust}/M_{\rm gas} = 0.01$. If it is molecular hydrogen we set the molecular gas opacity to 0.01 $\rm{cm^2 \, g^{-1}}$, and if $T \ge T_{\rm \hii}$ we set the opacity to zero. The last of these is a numerical convenience, because we have not implemented scattering or photoionization chemistry, and because the regions in our computation with $T>T_{\rm \hii}$ will contain so little mass they will be optically thin to the direct radiation field. We assume a dust-to-gas ratio of 0.01.

We solve equation (\ref{eqn:rtray}) by discretizing it along the line segments defined by the intersection of the ray with the cells of the computational mesh, considering only the most highly spatially resolved data at any given position. Specifically, when a ray with luminosity $L_{{\rm ray},j}$ passes through a cell along a segment of length $dl$, the optical depth of the segment is $\tau_j = \kappa_j\rho\, dl$ where $\rho$ is the dust (gas) density when the primary absorber is dust (molecular gas); and the luminosity of the ray decreases by an amount
\begin{equation}
dL_{{\rm ray},j} = L_{{\rm ray},j} \left(1 - e^{-\tau_j}\right).
\end{equation}
Here we compute $dl$ following the method of \citet{Wise2011a} as the ray transverses a cell. In the process, the cell absorbs an amount of energy and momentum at a rate
\begin{eqnarray}
\label{eqn:dedt}
\dot{\varepsilon}_{\rm rad, \, ray} & = & \sum_{j=1}^{N_\nu} dL_{{\rm ray},j} \\
\label{eqn:dpdt}
\dot{\mathbf{p}}_{\rm rad, \, ray} & = &  \sum_{j=1}^{N_\nu} \frac{dL_{{\rm ray},j}}{c} \mathbf{n}.
\end{eqnarray}
The total energy and momentum absorption rates for each cell,  $\dot{\varepsilon}_{\rm rad}$ and $\dot{\mathbf{p}}_{\rm rad}$, that are supplied to Equations (\ref{eqn:com}) and (\ref{eqn:coe}), are simply the sum of $\dot{\varepsilon}_{\rm rad, \, ray}$ and $\dot{\mathbf{p}}_{\rm rad, \rm ray}$ over all rays from all stars that pass through it, respectively. We terminate a ray when $L_{{\rm ray},j}(r) < 0.001 L_{{\rm ray},j}(0)$, i.e., when 99.9\% of the energy originally assigned to that ray has been absorbed, if it exits the computational domain, or has left the collapsing core. The last deletion criterion significantly reduces the cost of the ray tracing step if rays leave the core because the ambient medium will not absorb any energy or momentum from the rays, and deleting rays after they have traveled at least ten cells in the ambient medium without encountering core material therefore saves the need to continue following them through the remainder of the computational volume.

\subsection{Initial and Boundary Conditions}
\label{sec:ics}
Our initial setup for all runs is as follows. We begin with an isolated sphere of molecular gas and dust with mass $M_{\rm c}=150 \; \rm{M_{\rm \odot}}$, radius $R_{\rm c} = 0.1$ pc, temperature $T_{\rm c}=20$ K, and density profile $\rho \propto r^{-k_{\rm \rho}}$ with $k_{\rm \rho}=1.5$. The resulting surface density, $\Sigma=M_{\rm c}/(\pi R_{\rm c}^2) = 1 \; \rm{g \, cm^{-2}}$, is consistent with typical values observed in Galactic massive star forming regions \citep{McKee2003a, Swift2009a, Sanchez-Monge2013a, Tan2014a}. The resulting mean density of the core is $\bar{\rho}=2.4 \times 10^{-18} \;\rm{g \, cm^{-3}}$ ($1.2 \times 10^6 \; \rm{H \; nuclei \; cm^{-3}}$) and the characteristic free-fall collapse time scale is
\begin{equation}
t_{\rm ff} = \sqrt{\frac{3 \pi}{32 G \bar{\rho}}} \approx 42.6 \; \rm{kyr}.
\end{equation}
Our choice of $k_{\rm \rho}=1.5$ for the core density profile is in agreement with observations of star-forming regions at the $\sim$ 1 pc clump scale \citep{Caselli1995a, Mueller2002a, Beuther2007a} and the $\sim$0.1 pc scale \citep{Zhang2009a, Longmore2011a, Butler2012a, Stutz2016a}, which typically have $k_{\rm \rho}$ values within the range of 1.5-2.0.

Each core is placed at the center of a 0.4 pc box that is filled with a hot, diffuse ambient medium with a density equal to 1\% of the core edge material and a temperature of 2000 K so that the core is in thermal pressure equilibrium with its surroundings. We set the opacity of the ambient medium to zero so that the ambient gas is unable to cool. The base resolution for each run is $128^3$ and we allow for five levels of factors of two in refinement giving a maximum resolution of $4096^3$ cells on the finest level ($\Delta x_5$ = 20 AU).  We initially fill the entire domain with a blackbody radiation field equal to $E_{\rm 0} = 1.21 \times 10^{-9} \; \rm erg \; cm^{-3}$ corresponding to a 20 K blackbody. 

We consider two classes of initial condition: laminar cores and turbulent cores. For the laminar core we impose initial solid-body rotation at a rate such that the rotational energy of the core is 4\% of its gravitational binding energy (i.e., $E_{\rm rot}/|E_{\rm grav}|=0.04$). Our choice follows from the work of \citet{Goodman1993a}, which found that dense cores have values of  $E_{\rm rot}/|E_{\rm grav}|$ within $\sim 0.01-0.09$ with a typical value of 0.02 where the authors assumed the cores followed a uniform density profile. For the case of cores that follow a $\rho \propto r^{-3/2}$ density profile, like the cores simulated in this work, these values are reduced by a factor of two. We do not impose a net rotation for the turbulent core run and instead give the gas an initial weakly turbulent velocity field with a non-thermal one-dimensional velocity dispersion $\sigma_{\rm 1D}=0.4 \; \rm{km \; s^{-1}}$ corresponding to a virial ratio $\alpha_{\rm vir} \approx 5\sigma^2_{\rm 1D} R_{\rm c}/G M_{\rm c} = 0.12$ and we allow the turbulence in the core to decay freely, ensuring the core will undergo immediate collapse. The velocity power-spectrum imposed follows a Burger's turbulence spectrum, $P(k) \propto k^{-2}$, as is expected for supersonic turbulence \citep{Padoan1999a, Boldyrev2002a, Offner2009a}. We include modes between $k_{\rm min} =1$ and $k_{\rm max}=256$ and our turbulence mixture is chosen to be a mix of 2/3 solenoidal and 1/3 compressive modes, which is the natural mixture for a 3D fluid \citep{federrath2010a}. 

Our boundary conditions for the radiation, gravity, and hydrodynamic solvers are as follows. For each radiation update, we impose Marshak boundary conditions that bathe the simulation volume with radiation from a 20 K blackbody but allows radiation generated within the simulation volume to escape freely \citep{Krumholz2009a, Cunningham2011a, Myers2013a}. We set the gravitational potential, $\phi$, to zero at all boundaries when solving Equation (\ref{eqn:pois}) \citep{Myers2013a}. Finally, we impose outflow boundary conditions for the hydrodynamic update, meaning that we set the gradients of the hydrodynamic quantities ($\rho$, $\rho \mathbf{v}$, $\rho e$) to be zero at the domain when advancing the hyperbolic subsystem of equations \citep{Cunningham2011a,Myers2013a}.

We conduct four simulations. The first, which we call \noturb\ (where \texttt{RT} denotes that this simulation includes ray tracing), follows the collapse of a laminar pre-stellar core with the setup described above and includes our \harm\ hybrid radiation scheme to model the direct (with an adaptive raytracing scheme) and indirect radiation pressure (with FLD). Our second run, named \nort, is identical to \noturb\ except that it only includes the FLD approximation for the indirect radiation field and assumes that the stellar radiation energy is deposited close to the star. In this run we set the terms $\dot{\varepsilon}_{\rm rad}$ and $\dot{\mathbf{p}}_{\rm rad}$ to zero and add the source term $\sum_i L_{\it \star,i} W(\mathbf{x} - \mathbf{x}_i)$, where $L_{\it \star,i}$ is the combined accretion and stellar luminosity for star $i$, to Equation (\ref{eqn:coer}). This term simply adds the radiation energy injected by stars to the radiation energy density over the window kernel $W(\mathbf{x} - \mathbf{x}_i)$, which extends to a radius of four fine-level cells around each sink particle. We include this run to compare how the choice of the treatment of the radiation field can affect our results. The third run, named \noturblr, is a repeat of run \noturb\ but with a factor of 2 worse resolution ($\Delta x_{\rm min} = 40$ AU rather than 20 AU), and with significantly less stringent refinement criteria, as discussed in the next section. We include this run to determine how the results depend on the resolution. Our final run, which is called \turb, aims to be a better representation of massive star formation because star forming cores are turbulent, and this run follows the collapse of a turbulent pre-stellar core with the properties described above and includes our hybrid radiative transfer treatment. The initial numerical conditions for our simulations are summarized in Table \ref{tab:sim}. 

\begin{table*}
	\begin{center}
	\caption{
	\label{tab:sim}
Simulation Parameters
}

	\begin{tabular}{ l c  c  c  c  c  c  c c}
	\hline
	\textbf{Run} & & \nort\ & \noturb\ & \noturblr\ & \turb\ \\
	\\
	\hline
	\textbf{Physical Parameter}\\
	\hline
	\hline
	Cloud Mass [$\rm M_{\rm \odot}$] & $M_{\rm c}$ & 150 & 150 & 150 & 150\\
	Cloud Radius [pc] & $R_{\rm c}$ & 0.1 & 0.1 & 0.1 & 0.1 \\
	Surface Density [$\rm g \, cm^{-2}$] & $\Sigma$ & 1 & 1 & 1 & 1\\
	Temperature [K] & $T_{\rm c}$ & 20 & 20 & 20 & 20 \\
	Mean Density [$\rm 10^{-18}\, g \, cm^{-3}$] & $\bar{\rm \rho}_{\rm cl}$ & $2.4$ & $2.4$ & $2.4$ & $2.4$\\
	Mean Free-fall Time  [kyr] & $t_{\rm ff}$ & 42.6 & 42.6 & 42.6 & 42.6 \\
	Power Law Index & $k_{\rm \rho}$ & 1.5 & 1.5 & 1.5 &1.5\\
	Rotational Energy Ratio  & $E_{\rm rot}/|E_{\rm grav}|$ & 0.04 & 0.04 & 0.04 & --\\
	Velocity Dispersion [$\rm km \, s^{-1} $]& $\sigma_{\rm 1D}$ & -- & -- & -- & 0.4 \\
	\hline
	\\
	\textbf{Numerical Parameter}\\
	\hline
	\hline
	Rad. Trans. Scheme & & FLD & \harm\ & \harm\ & \harm\  \\
	Domain Length [pc] & $L_{\rm box}$ & 0.4 & 0.4 & 0.4 & 0.4 \\
	Base Grid Cells & $N_{\rm 0}$ & $128^3$ & $128^3$ & $128^3$ & $128^3$ \\ 
	Maximum Level & $l_{\rm max}$ & 5 & 5 & 4 & 5 \\
	Minimum Cell Size [AU] & $\Delta x_{\rm l_{\rm max}}$ &  20 & 20 & 40 & 20 \\
	Jeans Length Refinement &$J_{\rm max}$ & 0.125 & 0.125 & 0.125 & 0.125 \\
	$E_{\rm R}$ Gradient Refinement & $E_{\rm R}/\Delta x$ & 0.15 & 0.15 & -- & 0.15 \\  
	\hline
	\\
	\textbf{Simulation Outcome} \\
	\hline
	\hline
	Simulation Time [$t_{\rm ff}$] & &0.87 &0.70 & 0.70 & 0.87\\
	Massive Star Mass [$\rm M_{\odot}$] & &44.07 & 40.40 & 48.47 & 61.63\\
	Number of Sinks \tablenotemark{a} & &13 & 30 & 6 & 4 \\
	\end{tabular}
	\tablenotetext{a}{Final number of sinks with masses greater than 0.01 $M_{\rm \odot}$.}
	\end{center}
\end{table*}

\subsection{Refinement Criteria and Sink Creation}
\label{sec:sink}
The major advantage of AMR codes over fixed codes is that the user can adaptively refine on areas of interest. This is advantageous in astrophysical simulations, especially star formation simulations, which have large dynamic range but within which only certain regions of the domain require high resolution (e.g., high density regions in a molecular cloud that can undergo gravitational collapse to form stars). As the simulation evolves the AMR algorithm automatically adds and removes finer grids based on certain refinement criteria set by the user.

For each simulation we begin with a base grid with volume (0.4 pc)$^3$ discretized by $128^3$ cells and allow for five levels of refinement. This choice leads to a maximum resolution of 20 AU on the finest level. As the simulation evolves we continuously flag cells for refinement so that we can resolve areas in which stars may form or where instabilities may develop, such as gravitational and Rayleigh Taylor instabilities. In all simulations presented in this paper we flag cells for refinement if they meet one or more of the following criteria.
\begin{enumerate}
\item We refine any cell on the base level (i.e., level 0) that has a density equal to or greater than the core's edge density. This ensures that the entire pre-stellar core is refined to level 1 at the start of the simulation.
\item We refine any cell where the density in the cell exceeds the Jeans density given by
\begin{equation}
\label{eqn:rhoj}
\rho_{\rm max,J } = \frac{\pi J^2_{\rm max} c_{\rm s}^2}{G \Delta x_l^2}
\end{equation}
where $c_{\rm s}= \sqrt{k_B T/(\mu m_{\rm H})}$ is the isothermal sound speed, $\Delta x_l$ is the cell size on level $l$, and $J_{\rm max}$ is the maximum allowed number of Jeans lengths per cell \citep{Truelove1997a}. Throughout this work we take $J_{\rm max}=1/8$.
\item We refine any cell that is located within at least 8 cells of a sink particle.
\item We refine any cell within which the radiation energy density gradient exceeds
\begin{equation}
\label{eqn:egrad}
\left|\nabla E_{\rm R}\right| > 0.15 \frac{E_{\rm R}}{\Delta x_{\rm l}},
\end{equation}
i.e., where the radiation energy density changes by more than 15\% over the size of a single cell. This criterion is critical to ensuring that potentially-unstable interfaces are adequately resolved, and will become critical in our discussion later. Indeed, at late times in our simulations this criterion is responsible for refining more of the computational domain than any other one. We do not enforce this criterion for run \noturblr\ because this run aims to see how the development of RT instabilities depends on simulation resolution.
\end{enumerate}

This procedure is repeated recursively on all levels after every two level updates.  A sink particle can only be created when the Jeans density is violated on the finest level. When we check this criterion on the finest level we set $J_{\rm max}=1/4$ in Equation (\ref{eqn:rhoj})  following the artificial fragmentation tests of \citet{Truelove1997a}. If a cell is flagged on the finest level because it exceeds the Jeans density we place a sink particle in that cell whose mass is the excess matter in that cell. The new sink particle will evolve according to the equations in Section \ref{sec:numerics}.

\begin{figure*}
\centerline{\includegraphics[trim=0.2cm 0.2cm 0.2cm 0.2cm,clip,width=0.75\textwidth]{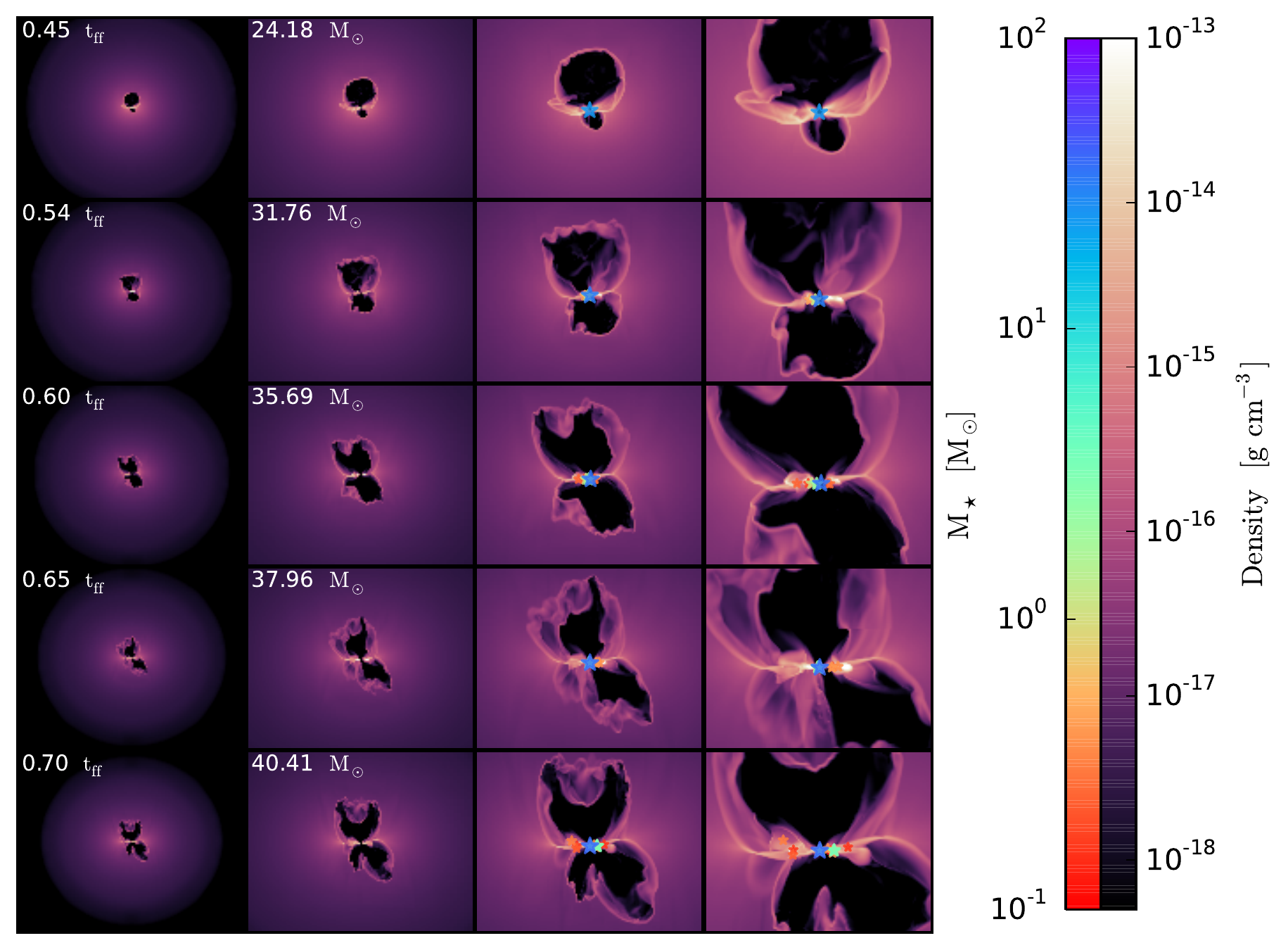}}
\caption{
\label{fig:noturbx}
A series of density slice plots along the $yz$-plane (edge-on views) showing the time evolution for our \noturb\ simulation. Each row corresponds to a specific snapshot where each panel is a zoom in of the previous panel by a factor of two from (40,000)$^2$ down to (5,000 AU)$^2$. The center of each panel corresponds to the center of the computational domain; stars with masses greater than $0.1 \, M_{\rm \odot}$ are over-plotted. The stars are color-coded by mass with the most massive being largest in size.  The time of the simulation and mass of the most massive star are given in the top-left corner of the first and second panels of each row, respectively.}
\end{figure*}

\subsection{Overall Algorithm}
We solve the equations described in Section \ref{sec:numerics} with the astrophysical AMR code \orion\ in a number of steps that we summarize below. First, we solve the equations of  hydrodynamics using a Godunov-type scheme with the HLLD approximate Riemann solver \citep{klein1999a, Miyoshi2005a}. Next, we incorporate self-gravity following the methods of \citet{truelove1998a} and \citet{klein1999a} by solving the Poisson equation (Equation (\ref{eqn:pois})) with an iterative multigrid scheme provided by the Chombo AMR Library \citep{Chombo}. In the third step we apply the \harm\ radiative transfer algorithm described in \citet{Rosen2016a}. The \harm\ update algorithm first applies an adaptive ray trace step for all star particles that belong to the computational domain to inject the radiation energy and momenta (Equations (\ref{eqn:dedt})-(\ref{eqn:dpdt})) from stars to the absorbing fluid and then performs the FLD step to evolve the radiation energy density and compute the radiation specific terms in Equations (\ref{eqn:cop})-(\ref{eqn:coer}) \citep{Rosen2016a}. The FLD step uses an operator split approach that first solves the radiation pressure, work, and advection terms explicitly, and then implicitly updates the gas and radiation energy densities for terms that involve diffusion and the emission/absorption of radiation \citep{Krumholz2007a}. The implicit solve update is handled by the iterative process described in \citet{Shestakov2008a} that uses pseudo-transient continuation to reduce the number of iterations. Finally, we update the sink particle states with Equations (\ref{eqn:mdot})-(\ref{eqn:dpi}) by computing their interactions with the fluid.

\section{Results}
\label{sec:results}
In this section we describe the results from our simulations presented in Section \ref{sec:ics} and summarized in Table \ref{tab:sim}. In Section \ref{sec:laminar} we first discuss our results for our laminar core run \noturb, which includes our new hybrid radiation transfer scheme. We then compare this simulation to run \nort, which only includes the radiation pressure associated with the diffuse dust-reprocessed radiation field. We defer discussion of our comparison low-resolution run, \noturblr\ to Section \ref{sec:rrt}. Next in Section \ref{sec:turb} we discuss our results from run \turb, which simulates the collapse of an initially turbulent core with our \harm\ algorithm. All simulations presented here were run on the NASA supercomputer Pleiades located at NASA Ames or the Hyades supercomputer located at UCSC. We run each simulation to the point where the timestep either becomes too short to be practical (as in the case of run \nort) or until the point that the simulation takes too long to evolve because the majority of the bubble shells are refined to the finest level, severely increasing the computational cost of the simulation (as in runs \noturb\ and \turb). We use the \texttt{yt} package \citep{Turk2011a} to produce all the figures and quantitative analysis shown below.

\subsection{Collapse of Laminar Pre-stellar Cores}
\label{sec:laminar}
Here we present the results of run \noturb. At the end of this simulation the most massive star has a mass of $40.40\; M_{\odot}$. We ran the simulation for a time of $t=0.70 \; t_{\rm ff}$.

\subsubsection{Evolution of Radiation Pressure Dominated Bubbles}
\label{sec:noturbbubble}
We show a series of density slices at various times for run \noturb\ in Figure \ref{fig:noturbx}. We find that a radiation pressure dominated bubble begins to expand in the polar direction above the star, but not below, when the star has reached a mass of $\sim 14.5 \; M_{\odot}$ at time $t=0.34 \; t_{\rm ff}$ (not shown). A radiation pressure dominated bubble only begins to expand below the star when it reaches a mass of $\sim 22.3 \; M_{\odot}$ at time $t=0.43 \; t_{\rm ff}$ (not shown).  As the radiation pressure dominated bubbles continue to expand small-scale RT instabilities begin to develop in the dense shells, but their growth is slow initially. This is likely due to the fact that the radiation pressure is able to push back on these instabilities, inhibiting their non-linear growth when the shell is optically thick to the direct stellar radiation field. For example, the absorption of the high-energy stellar radiation field is not fully resolved because the minimum optical depth through a 20 AU cell (the resolution on the finest level) is of order unity, where we have assumed that the shell density is $\rho \sim 10^{-16} \; \rm{cm^{-3}}$ and the dust opacity to the high energy radiation is $\kappa_{\rm UV} \sim 10^4 \; \rm{cm^2 \, g^{-1}}$ (e.g., see Figure \ref{fig:opacity}). 

When the primary star has a mass of $\sim 26.3 \; M_{\odot}$ at time $t=0.48 \; t_{\rm ff}$ the right side of the disk becomes flared and material is blown off the accretion disk by the direct radiation pressure. This injection of material into the upper bubble leads to an asymmetric absorption of the direct radiation field. This asymmetry and the resulting shielding of the direct radiation field causes RT instabilities to grow faster on the right side of the bubble shell while suppressing the non-linear growth of RT instabilities on the left side. Our results demonstrate that the seeding of RT instabilities and their resulting non-linear growth is a direct result of the asymmetric absorption of the energy and momentum from the direct radiation field across the bubble shell. Regions that feel a weaker radiative force are thus more likely to allow RT instabilities to grow non-linearly, leading to asymmetries in the bubble shells. These instabilities continue to grow as the simulation evolves. Once these unstable regions grow large enough gas is able to collapse directly onto the star-disk system if the material becomes sub-Eddington, delivering mass to the accreting protostar. We first see this behavior begin when the star reaches  $\sim 30 \; M_{\odot}$ at time $t=0.51 \; t_{\rm ff}$. At this time RT instabilities that develop on the right side of the bubble shell become sub-Eddington and grow large enough that the right edge of the shell deposits material onto the disk.

\begin{figure}
\centerline{\includegraphics[trim=0.2cm 0.2cm 0.2cm 0.2cm,clip,width=0.95\columnwidth]{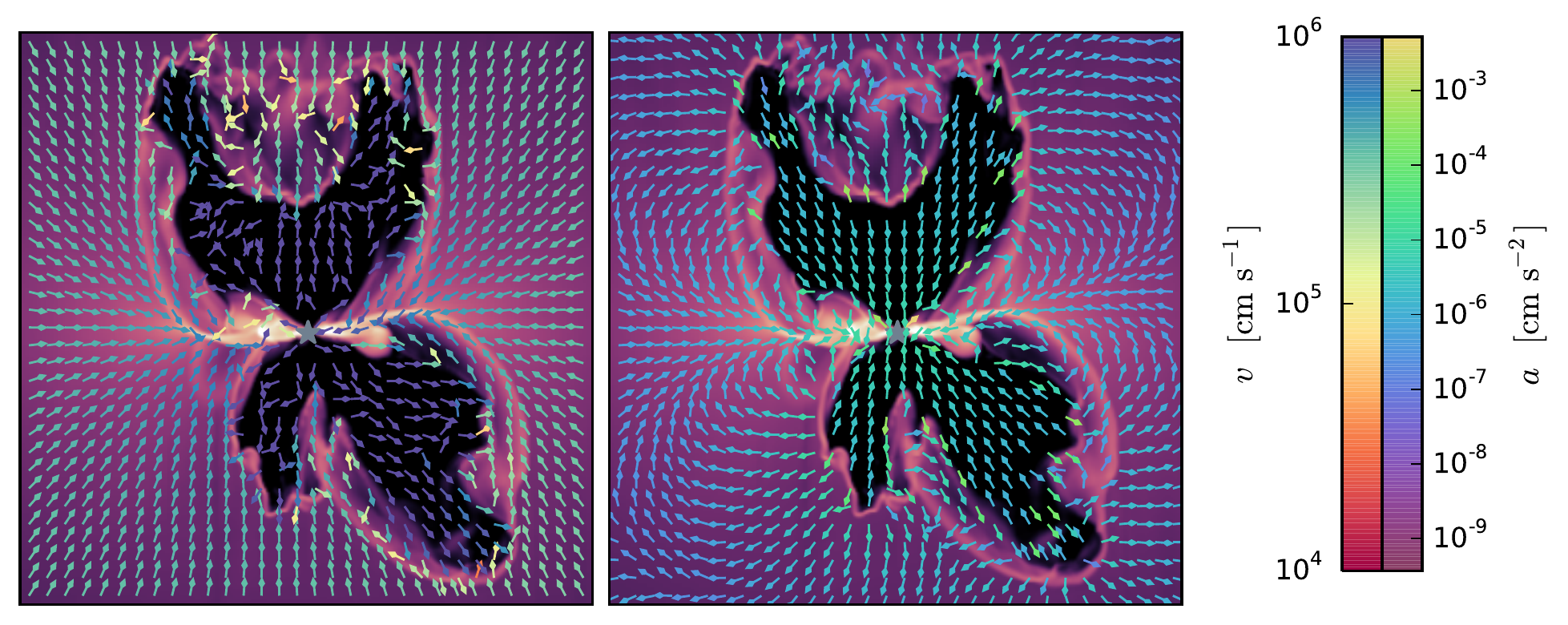}}
\caption{
\label{fig:accel}
Density slices along the $yz$-plane with the velocity field (left panel) and net acceleration due to gravity and radiation (right panel) over-plotted when the most massive star is 40.4 $M_{\rm \odot}$ at $t=0.7 \, t_{\rm ff}$ for run \noturb. The region plotted is (8,000 AU)$^{2}$ and the center of each panel corresponds to the center of the computational domain. The gray star denotes the position of the most massive star.
}
\end{figure}

As the simulation progresses RT instabilities continue to develop across the top and bottom shells at an accelerating rate. This is a result of the star's movement in the disk, which causes the disk to shadow the direct radiation field (see Section \ref{sec:noturbdisk}) and is also due to the increasing surface density of the bubble shells as the core collapses. The bottom shell goes unstable and begins to collapse when the star is $\sim 34.3 \; M_{\odot}$ at time $t=0.57 \; t_{\rm ff}$ and this material reaches the disk at time $t=0.62 \; t_{\rm ff}$. Material from the collapsed left side of the bottom bubble shell continues to deliver mass to the star-disk system until the star has reached a mass of $\sim 38.6 \; M_{\odot}$ at time $t=0.66 \; t_{\rm ff}$. At this point the direct radiation pressure causes the left side of the bottom bubble to expand again. 

\begin{figure}
\centerline{\includegraphics[trim=0.2cm 0.2cm 0.2cm 0.2cm,clip,width=0.9\columnwidth]{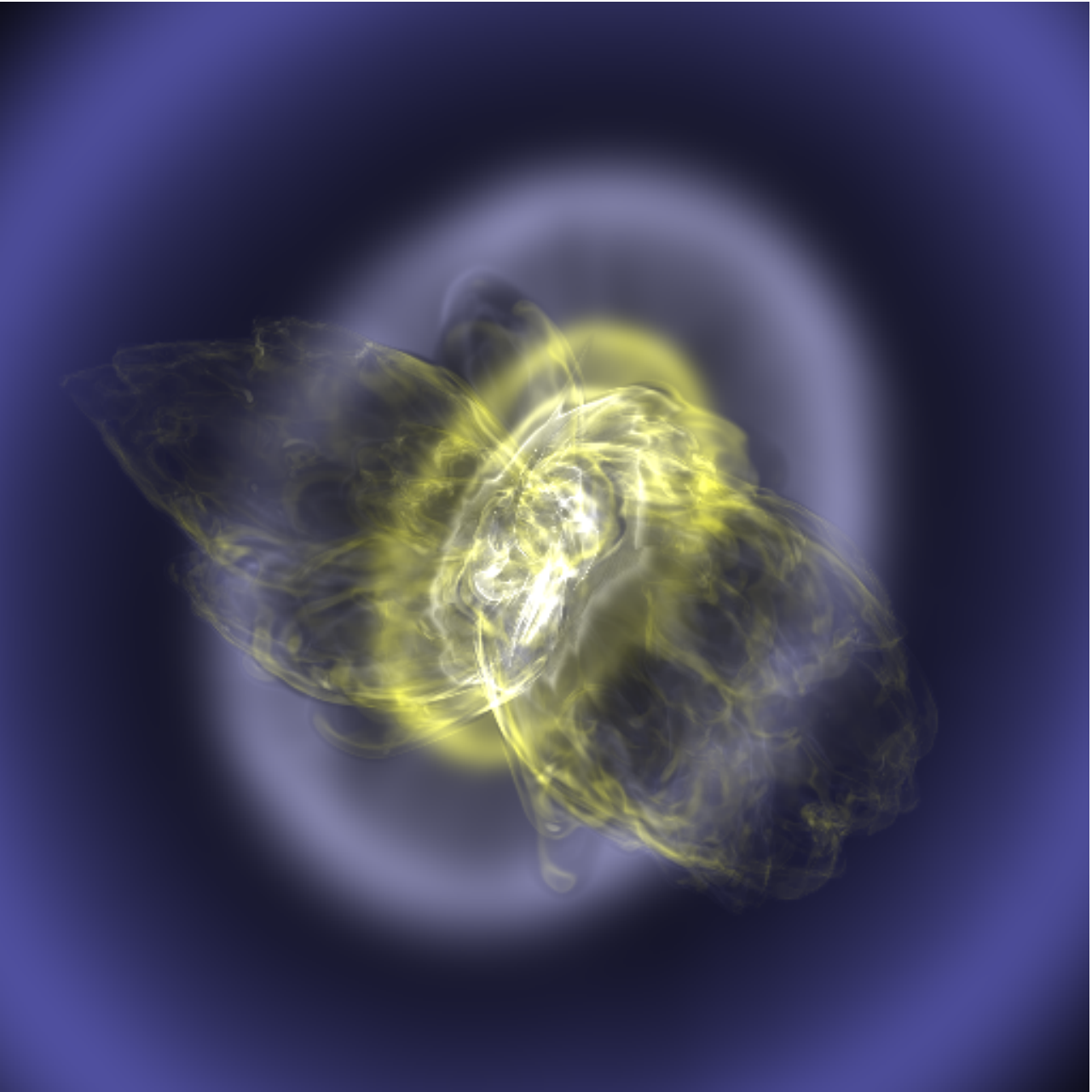}}
\caption{
\label{fig:vol}
Volume rendering of a snapshot from run \noturb\ when the star is 40.1 $M_{\rm \star}$ at time $t=0.69 \; t_{\rm ff}$ that shows RT instabilities are common throughout the radiation-pressure dominated bubbles.
}
\end{figure}

At the end of the simulation we see that regions of the top and bottom bubbles are collapsing towards the star (bottom panel of Figure \ref{fig:noturbx}). To demonstrate this, we show the velocity (left panel) and net acceleration of the gas due to radiative and gravitational forces (right panel) in Figure \ref{fig:accel} at the end of run \noturb. This Figure shows that the gas velocities in some of the densest regions of the shell are in the direction of the star-disk system even though the net acceleration along the majority of the bubble shells tend to point away from the star. The regions that experience a weaker net acceleration are more likely to go unstable while the regions that feel a larger net force will expand away from the star at a faster rate leading to more asymmetries in the bubble shells. It is these regions that go unstable and grow non-linearly, allowing material to continue to fall towards the star. This process may ultimately supply mass to the star-disk system and be accreted onto the star. Figure \ref{fig:vol} shows that the majority of the bubble shells have become RT unstable. 

Run \noturb\ also shows that throughout the collapse of the laminar core and growth of the massive protostar, a considerable amount of material is delivered from the edges of the radiation pressure dominated bubbles to the accretion disk via RT instabilities because these regions are shielded from the direct radiation field. We find that shielding of the direct radiation field promotes RT instabilities because the asymmetric absorption of the direct radiation field, which causes the direct radiation force to vary over the inner surface of the bubble shells, can lead to perturbations that will then amplify and become RT unstable. This can be seen by observing where the stellar radiation field is absorbed in the bubble shells. The left panel in Figure \ref{fig:frad} shows a zoomed-in density slice plot of the star at the end of the simulation  along the $yz$-plane. Vectors showing the direction and magnitude of the direct radiation momentum deposition are over-plotted. The right panel shows the acceleration from the diffuse dust-reprocessed radiation field. Stellar radiation is able to stream freely along the polar directions that are not shielded by the accretion disk and gas within the bubble. In contrast, the accretion disk shields part of the radiation field near the left and right sides of the star. Furthermore, the radiation is shielded to a greater degree on the left side of the star because the disk is flared. Indeed, Figure \ref{fig:frad} shows that the left side of the top and bottom bubbles experience a greater degree of instability than the right side of the bubbles suggesting that the growth and subsequent collapse of these RT instabilities depends on the shielding and resulting patchiness of the direct radiation field.

We find that throughout the simulation regions of the bubble shells that are shielded by the accretion disk feel a weaker direct radiative force and are more likely to go unstable and bring material to the star-disk system. This can be seen in Figure \ref{fig:feddnoturb} that shows the same snapshot with the acceleration vectors over-plotted. In this Figure, the color of each vector is the value for the Eddington ratio, $f_{\rm edd}= |\mathbf{f}_{\rm rad}|/|\mathbf{f}_{\rm grav}|$, where we have included the contribution from both the direct and dust-reprocessed radiation fields. Values of $f_{\rm edd} \lesssim 1$ are subject to collapse. The bubble interiors have $f_{\rm edd} \ll 1$ because the bubble interiors are optically thin whereas regions of the bubble shells that become unstable have $f_{\rm edd} \lesssim 1$. Our results demonstrate that RT instabilities, along with disk accretion, deliver mass to the star, and that these instabilities become more important as the system evolves.

\subsubsection{Accretion Disk Evolution}
\label{sec:noturbdisk}

\begin{figure*}
\centerline{\includegraphics[trim=0.2cm 0.2cm 0.2cm 0.2cm,clip,width=0.75\textwidth]{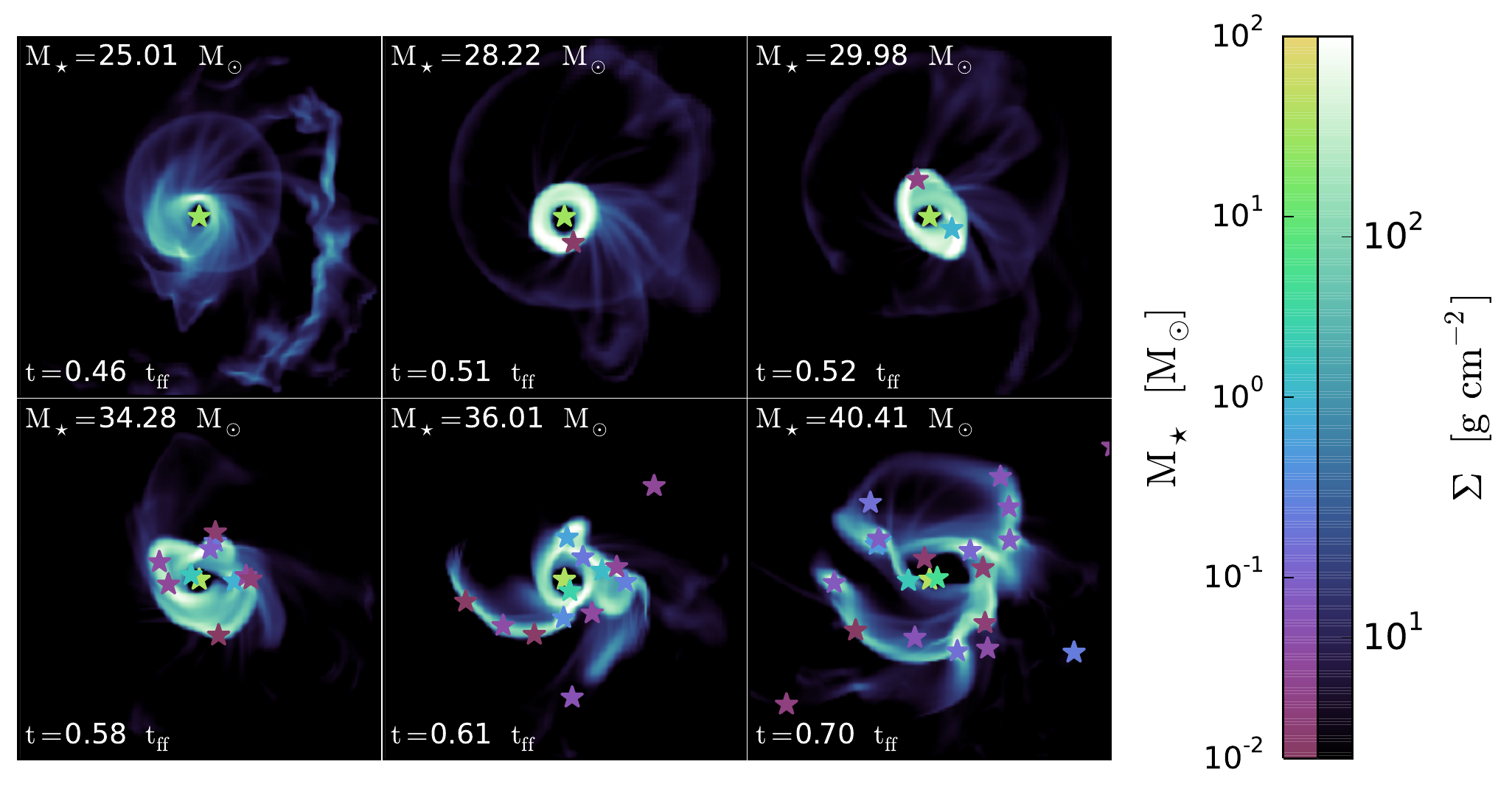}}
\caption{
\label{fig:noturbz}
Surface density projections of the accretion disk in run \noturb\ showing the disk's time evolution. Each panel represents a projection of the accretion disk, with the most massive star at the center of the panel, that is (3000 AU)$^2$ in size. The projection is taken over a height of 1000 AU above and below the massive star. Stars with masses greater than 0.01 $M_{\rm \odot}$ are over-plotted on all panels.
}
\end{figure*}

Next we examine the behavior and growth of the accretion disk. Figure \ref{fig:noturbz} shows a series of surface density plots along the plane perpendicular to the core's rotation axis ($xy$-plane) that show the growth and evolution of the accretion disk around the massive star. The top left hand panel of Figure \ref{fig:noturbz} shows that a noticeable thick accretion disk begins to form when the star reaches  $\sim 25 \; M_{\odot}$ (i.e., an accretion disk with a radius larger than the 80 AU accretion zone radius of the sink particle). The accretion disk continues to grow in size as the core undergoes inside-out collapse because material with a higher starting radius and thus a larger specific angular momentum is circularized farther away from the star. As the disk evolves it develops spiral arms that become unstable and fragment into companions. The first companion star forms at time $t=0.51 \; t_{\rm ff}$ when the primary star has a mass of $\sim 28.2 \; M_{\odot}$. The combined interaction of the primary star, accretion disk, and companions induces gravitational torques leading to even more companions. By the end of the run the primary has a mass of $\sim 40.4 \; M_{\odot}$ and has 29 companion stars with masses greater than 0.01 $M_{\rm \odot}$. The most massive companion is only $4.43 \; M_{\odot}$; sixteen of the companion stars have masses greater than $0.1 \; M_{\odot}$, but only four of these have masses greater than $1 \; M_{\odot}$. Thus at the end of run \noturb\ we do not have a massive binary system, but rather a hierarchical system consisting of a massive primary and a series of much less massive mass companions. Figure \ref{fig:mdot} shows the total growth in mass of the primary star and its companions as a function of time (top panel).

\subsubsection{Comparison to Run \nort}
\label{sec:comp}

To determine how the results depend on our treatment of the direct radiation field, we perform run \nort, a comparison run that does not include the direct radiation field and instead deposits the stellar radiation near the star. This method does not properly model the momentum deposition by the stellar radiation field and only includes gray dust opacities, which underestimates the true optical depth associated with the stellar radiation field. For example, the frequency dependent dust opacities range from $\sim10-1000 \; \rm{cm^2/g}$ for molecular gas (i.e., assuming a dust-to-gas fraction of 0.01) for the high-frequency stellar radiation (e.g., see Figure \ref{fig:opacity}) while the \citet{semenov2003a} opacities used for the FLD method in \orion\ range from $\sim1-10 \; \rm{cm^2/g}$ for molecular gas at temperatures below $T\lesssim 1500$ K. Run \nort\ follows the same initial conditions as run \noturb\ but does not include the adaptive ray tracing from the \harm\ algorithm. Figures \ref{fig:nortx} and \ref{fig:nortz} show the time evolution for run \nort, and are analogous to Figures \ref{fig:noturbx} and \ref{fig:noturbz} for run \noturb. 

In run \nort\ the radiation pressure dominated bubbles begin to expand along the polar directions (both above and below the star) when the star reaches $\sim18 \; M_{\odot}$ (not shown in Figure \ref{fig:nortx}) whereas in run \noturb\ a radiation-pressure driven bubble began to expand above (below) the star when it reached a mass of $\sim$15 $M_{\rm \odot}$ ($\sim$22 $M_{\rm \odot}$). Similarly, \citet{Kuiper2012a} also found that their massive star formation simulation, which only included FLD, launched radiation driven bubbles earlier than their comparison run that included both ray tracing and FLD. Comparison of Figures \ref{fig:noturbx} and \ref{fig:nortx} also shows that direct radiation pressure is more efficient at evacuating material interior to the bubble walls while also causing substantial RT instabilities to begin to develop later. This is also demonstrated in the top panel of Figure \ref{fig:fradcone}, which shows the volume weighted mass density as a function of radial distance of a three-dimensional cone above the center of the computational domain. In run \nort, the bottom bubble becomes unstable and collapses onto the disk when the star has reached a mass of $\sim$23.7 $M_{\rm \odot}$ while the bottom bubble first becomes unstable in run \noturb\ and begins to collapse when the star has a mass of $\sim35 \; M_{\rm \odot}$. This difference is due to the fact that the direct radiation force falls off as $r^{-2}$ so infalling material will feel a greater force as it falls towards the star, causing the direct radiation to push the material back towards the shell; whereas the diffuse radiation pressure is roughly constant in the bubbles because it follows the radiation energy density. Therefore, the diffuse radiation pressure is less likely to inhibit the non-linear growth of RT instabilities allowing the shells to collapse earlier. As the star becomes more luminous in both runs the bottom bubbles re-expand. However, we find that, regardless of the radiation treatment, the radiation dominated bubbles eventually become unstable and deliver mass to the star-disk system through RT instabilities.

In agreement with \citet{Kuiper2012a} we also find that neglecting the direct radiation field leads to underestimating the true radiation force density. Figure \ref{fig:fradcomp} shows volume weighted projection plots of the direct (top left panel), diffuse (top right panel), and total radiation force densities (bottom left panel) for a snapshot of run \noturb\ when the primary star has a mass of $36.1\; M_{\rm \odot}$. The bottom right panel of Figure \ref{fig:fradcomp} shows the total radiation force density for run \nort\ at the same stellar mass for comparison.  The top two panels shows that the radiation force density associated with the direct radiation field is much greater than the diffuse component in regions of the bubble shells where the direct component is absorbed while comparison of the bottom two panels demonstrate that the radiation force density is greatly underestimated at the location of the bubble shells where the direct radiation is absorbed. This is also observed in the bottom panel of Figure \ref{fig:fradcone} which shows the volume weighted averaged  direct, diffuse, and total radiation force densities as a function of radius for a three-dimensional cone above the center of the computational domain for the snapshots shown in Figure \ref{fig:fradcomp}. We also find the integrated radiative force over a spherical volume with radius 7000 AU around the primary star in run \noturb\ is a factor of $\sim2$ larger than run \nort\ when the star is $36.1\; M_{\rm \odot}$. Thus, we find that inclusion of the direct radiation field leads to a larger total radiation force as expected but regions of the bubble shells still become RT unstable regardless.

Although the development of RT instabilities is qualitatively the same for runs \noturb\ and \nort, the structure of the accretion disk and the consequent creation of companions is not. Comparison of Figures \ref{fig:noturbz} and \ref{fig:nortz} show that the accretion disk in run \nort\ is more extended and has an overall lower surface density than in run \noturb. It also undergoes less fragmentation resulting in fewer companions. For example, the primary star in run \nort\ has eight companion stars with masses greater than 0.01 $M_{\rm \odot}$ when the primary has a mass of 40.4 $M_{\rm \odot}$, whereas the primary star in run \noturb\ has 29 companion stars when the primary has a mass of 40.4 $M_{\rm \odot}$. Furthermore, the most massive companion in run \nort\ is 11.28 $M_{\rm \odot}$ when the primary has a mass of 40.4 $M_{\rm \odot}$, a factor of $\sim$2.5 larger than the most massive companion in run \noturb\ for the same primary stellar mass.

Figure \ref{fig:mdot} shows the evolution of the primary and total companion star mass for both runs (top panel) and accretion rate onto the primary star (bottom panel). We find that although the number of companion stars formed is different for each run, the total mass contained in the companion stars is qualitatively similar when the primary star has a mass of 40.4 $M_{\rm \odot}$ (i.e., the final stellar mass in run \noturb). The accretion rate onto the primary star in each run is also qualitatively similar. Initially the accretion rate onto the primary star is smooth for both simulations, but it becomes chaotic once the disk becomes gravitationally unstable and forms companion stars. This chaotic behavior can be attributed to disk gravitational instabilities and RT instabilities funneling material to the stars. Since run \noturb\ undergoes a greater degree of disk fragmentation, we find that the total stellar mass in run \noturb\ is larger than that of run \nort\ at $t=0.70 \; t_{\rm ff}$ (the final time in run \noturb). The total stellar mass at this time is 55.80 and 49.68 $M_{\rm \odot}$ for run \noturb\ and run \nort, respectively.
The decrease in disk fragmentation in run \nort\ can be understood by looking at the temperature structure of the accretion disks as shown in Figure \ref{fig:tempdisk}. Collapse can only occur in regions that become Jeans unstable, and this instability depends on both the density and temperature of the gas. A hotter, lower density region is less likely to fragment (e.g., see Equation (\ref{eqn:rhoj})). Run \nort\  has a hotter accretion disk because the radiation is deposited in the immediate vicinity of the stars, and it subsequently diffuses through the disk, heating up the gas as shown in the bottom right panel of Figure \ref{fig:fradcomp}. In contrast, the absorption of radiation for run \noturb\ is highly dependent on the frequency-dependent optical depth of the cells that the rays transverse. Once an evacuated region appears above and below the disk, much of the stellar radiation energy is deposited in the bubble walls at a considerable distance from the disk, allowing the disk to remain cooler. Furthermore, the accretion rate onto the stars depends on the disk temperature, $\dot{M} \propto c_{\rm s}^3$ \citep{Shu1977a}. Thus the hotter gas in the accretion disk in run \nort\ produces more massive companion stars, consistent with what we observe.

\begin{figure}
\centerline{\includegraphics[trim=0.2cm 0.2cm 0.2cm 0.2cm,clip,width=0.9\columnwidth]{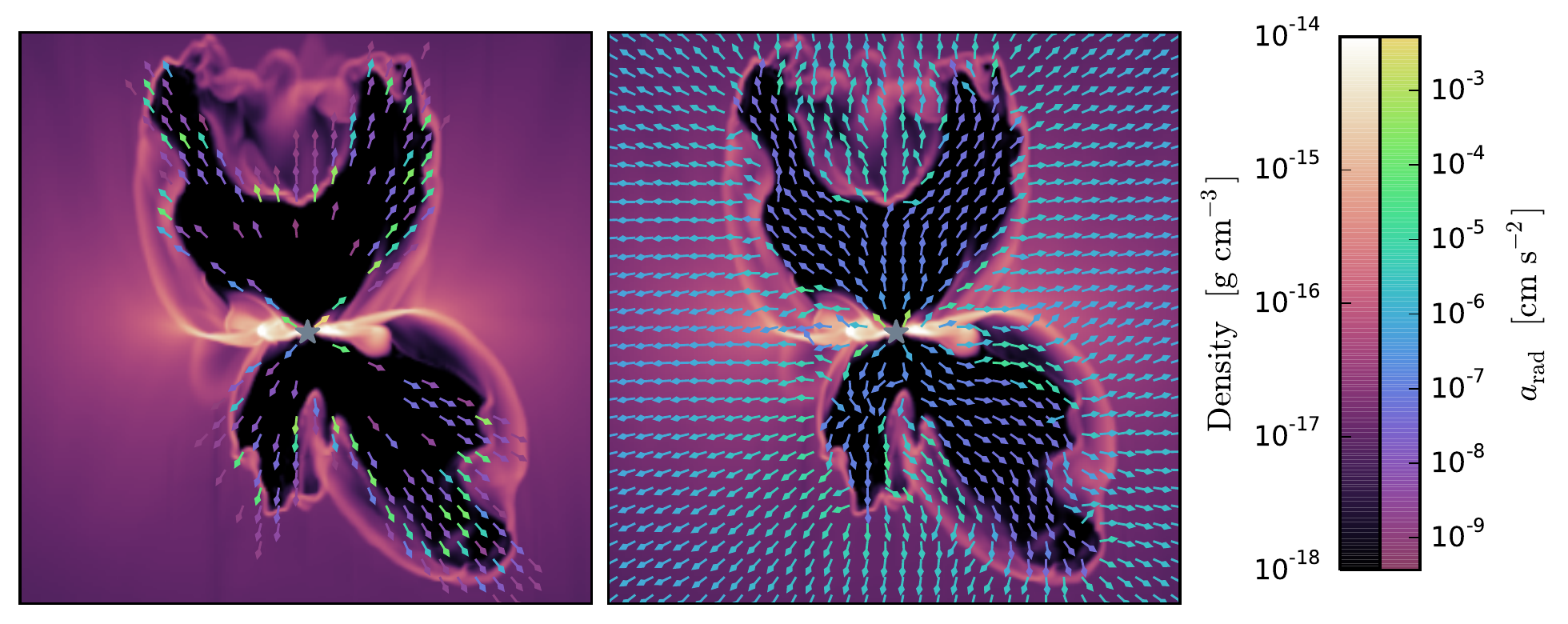}}
\caption{
\label{fig:frad}
Same as Figure \ref{fig:accel}, but now the vectors show the acceleration due to the direct (left panel) and diffuse (right panel) radiation fields.
}
\end{figure}

\begin{figure}
\centerline{\includegraphics[trim=0.2cm 0.2cm 0.2cm 0.2cm,clip,width=0.9\columnwidth]{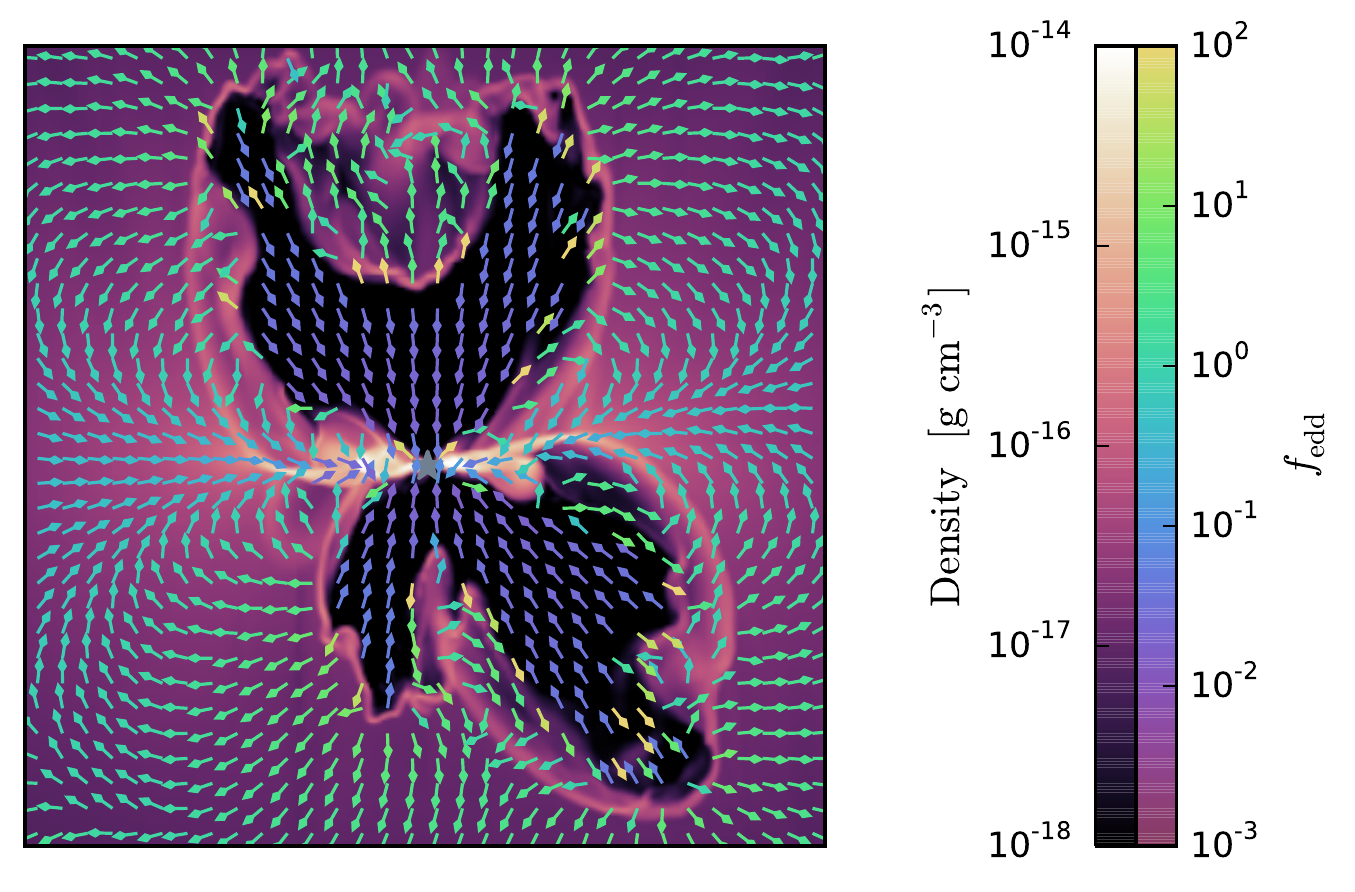}}
\caption{
\label{fig:feddnoturb}
Same as Figure \ref{fig:accel}, but here arrows show the direction of the net (gravitational plus radiative) acceleration. Vector colors show the Eddington ratio, $f_{\rm edd}=|\mathbf{f}_{\rm rad}|/|\mathbf{f}_{\rm grav}|$, where $\mathbf{f}_{\rm rad}$ is the total radiative force due to both the direct and diffuse components.
}
\end{figure}

\begin{figure}
\centerline{\includegraphics[trim=0.2cm 0.2cm 0.2cm 0.2cm,clip,width=0.8\columnwidth]{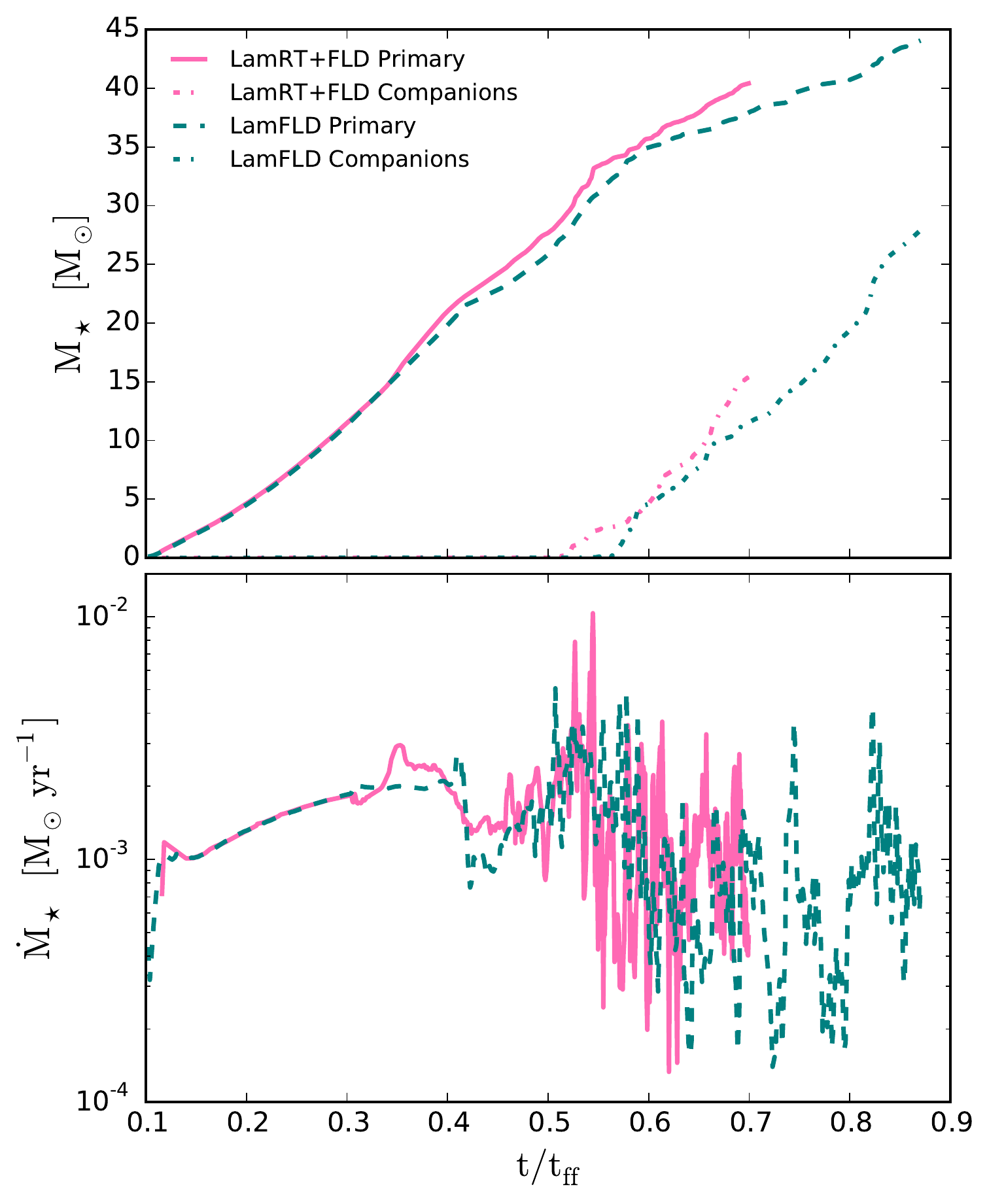}}
\caption{
\label{fig:mdot}
Stellar mass and accretion rates for runs \noturb\ and \nort. Top panel: Total mass in primary and companion stars as a function of time for run \noturb\ (pink solid and dot-dashed lines, respectively) and run \nort\ (teal dotted and dot-dashed lines, respectively). Bottom panel: Primary star accretion rate as a function of time for run \noturb\ (pink solid line) and run \nort\ (teal dashed line).
}
\end{figure}
 \begin{figure*}
\centerline{\includegraphics[trim=0.2cm 0.2cm 0.2cm 0.2cm,clip,width=0.7\textwidth]{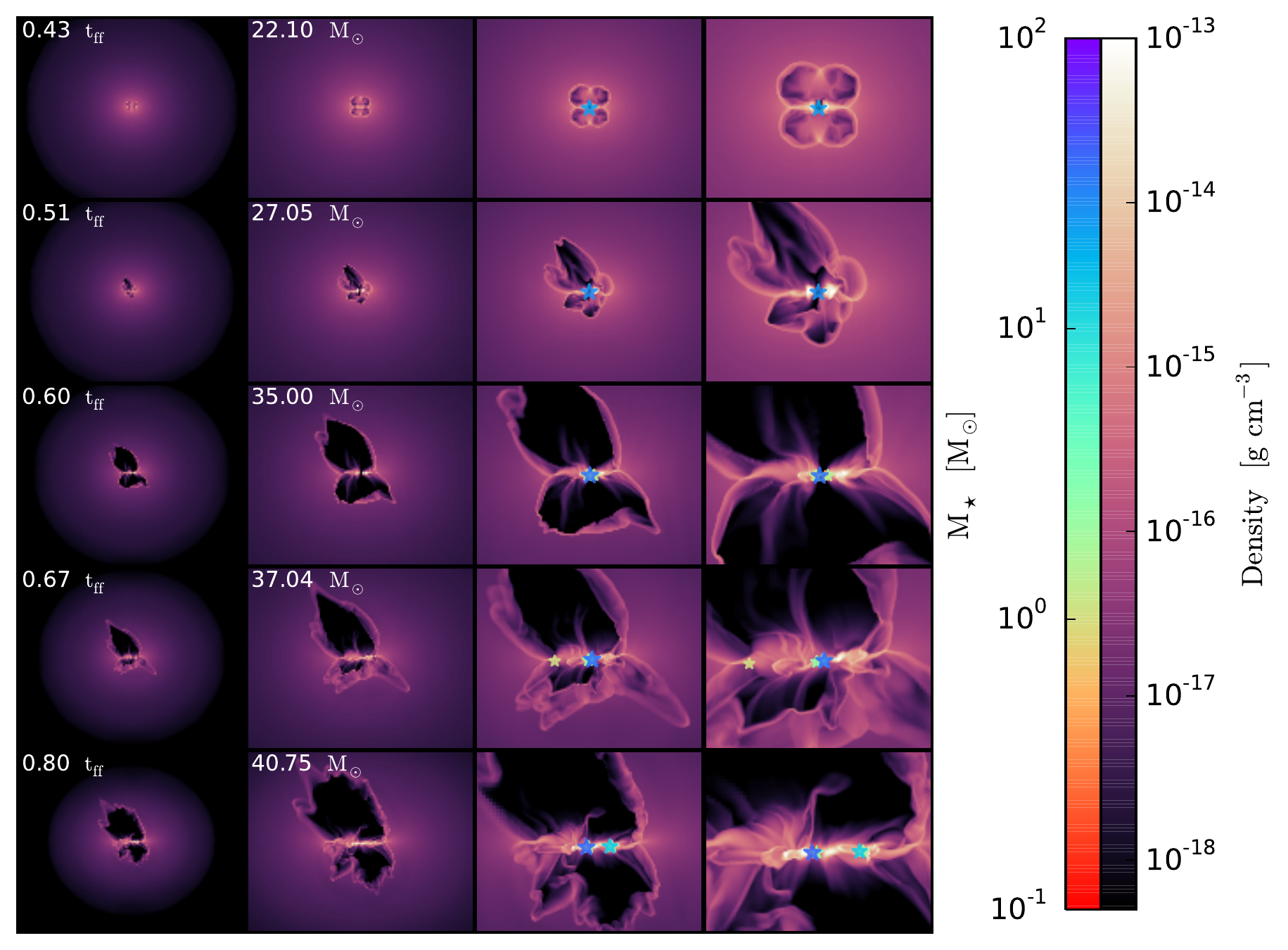}}
\caption{
\label{fig:nortx}
Same as figure \ref{fig:noturbx}, but for run \nort.
}
\end{figure*}

\begin{figure*}
\centerline{\includegraphics[trim=0.2cm 0.2cm 0.2cm 0.2cm,clip,width=0.7\textwidth]{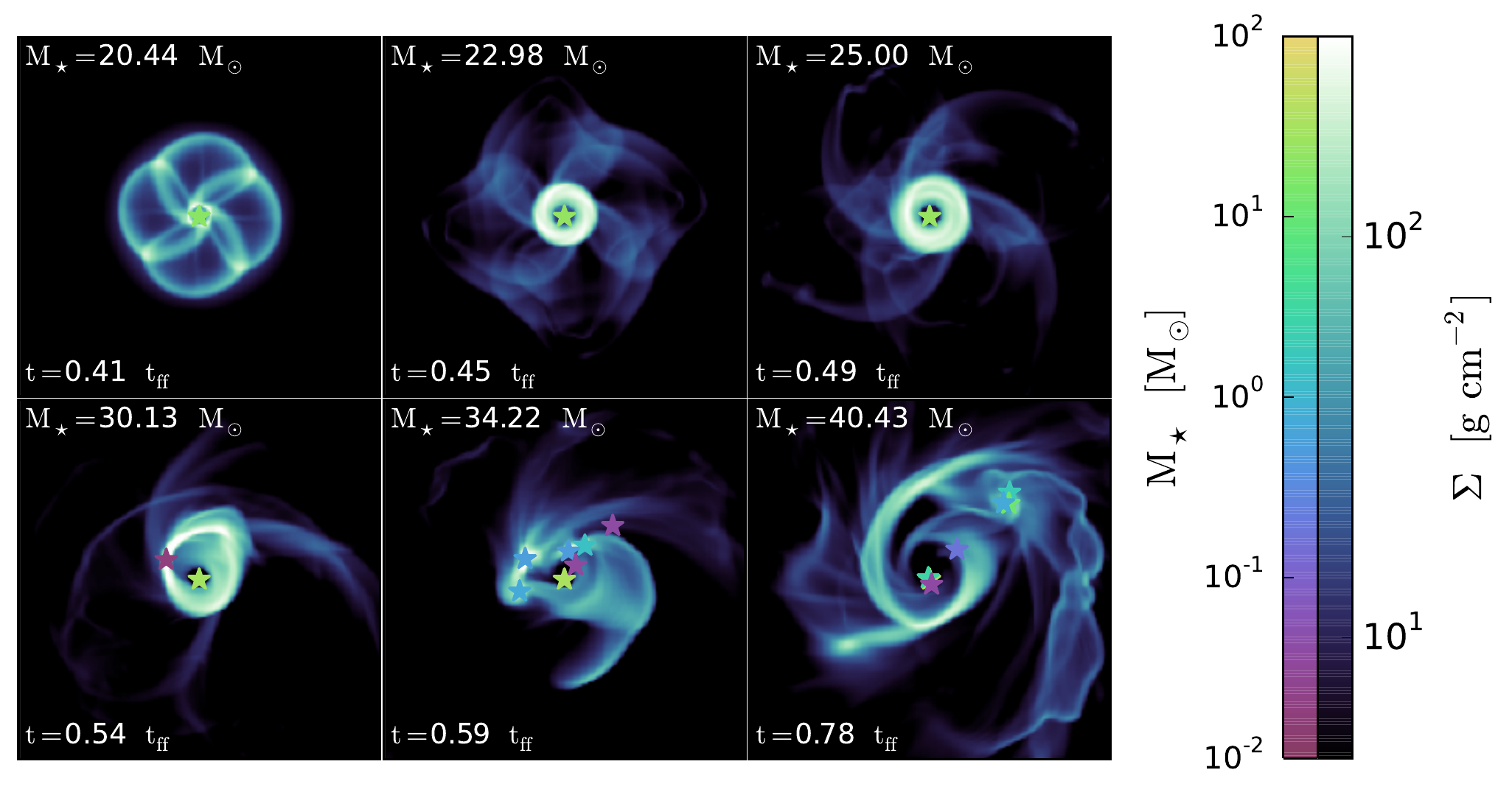}}
\caption{
\label{fig:nortz}
Same as figure \ref{fig:noturbz}, but for run \nort.
}
\end{figure*}

\begin{figure}
\centerline{\includegraphics[trim=0.0cm 0.0cm 0.0cm 0.0cm,clip,width=0.8\columnwidth]{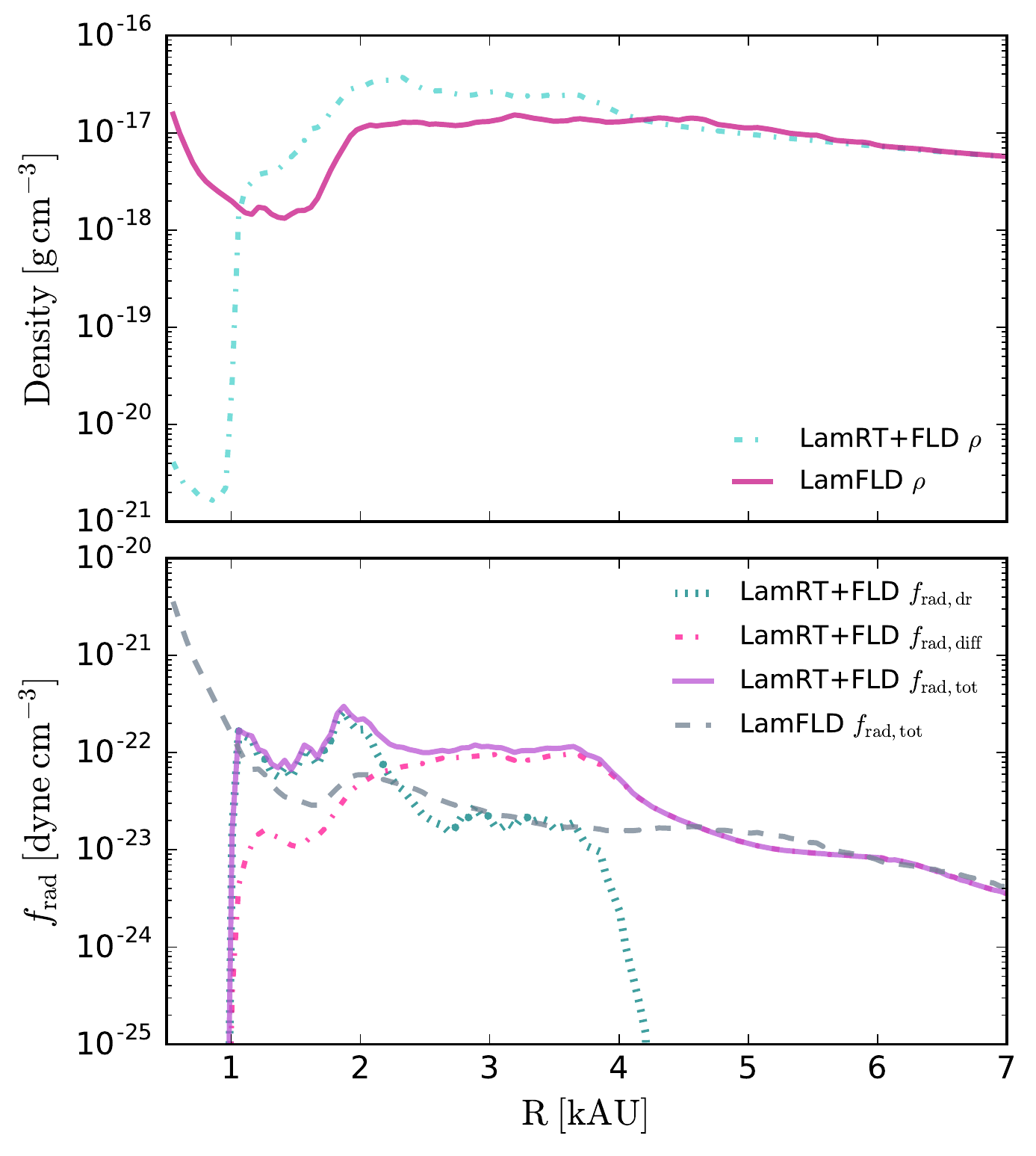}}
\caption{
\label{fig:fradcone}
Volume-weighted averaged mass densities (top panel) and direct, diffuse, and total radiation force densities (bottom panel)  as a function of radius for a three-dimensional cone above the center of the computational domain for runs \noturb\ and \nort\ when the primary star has a mass of 36.1 $M_{\rm \odot}$.
}
\end{figure}

\begin{figure}
\centerline{\includegraphics[trim=0.02cm 0.02cm 0.02cm 0.02cm,clip,width=0.9\columnwidth]{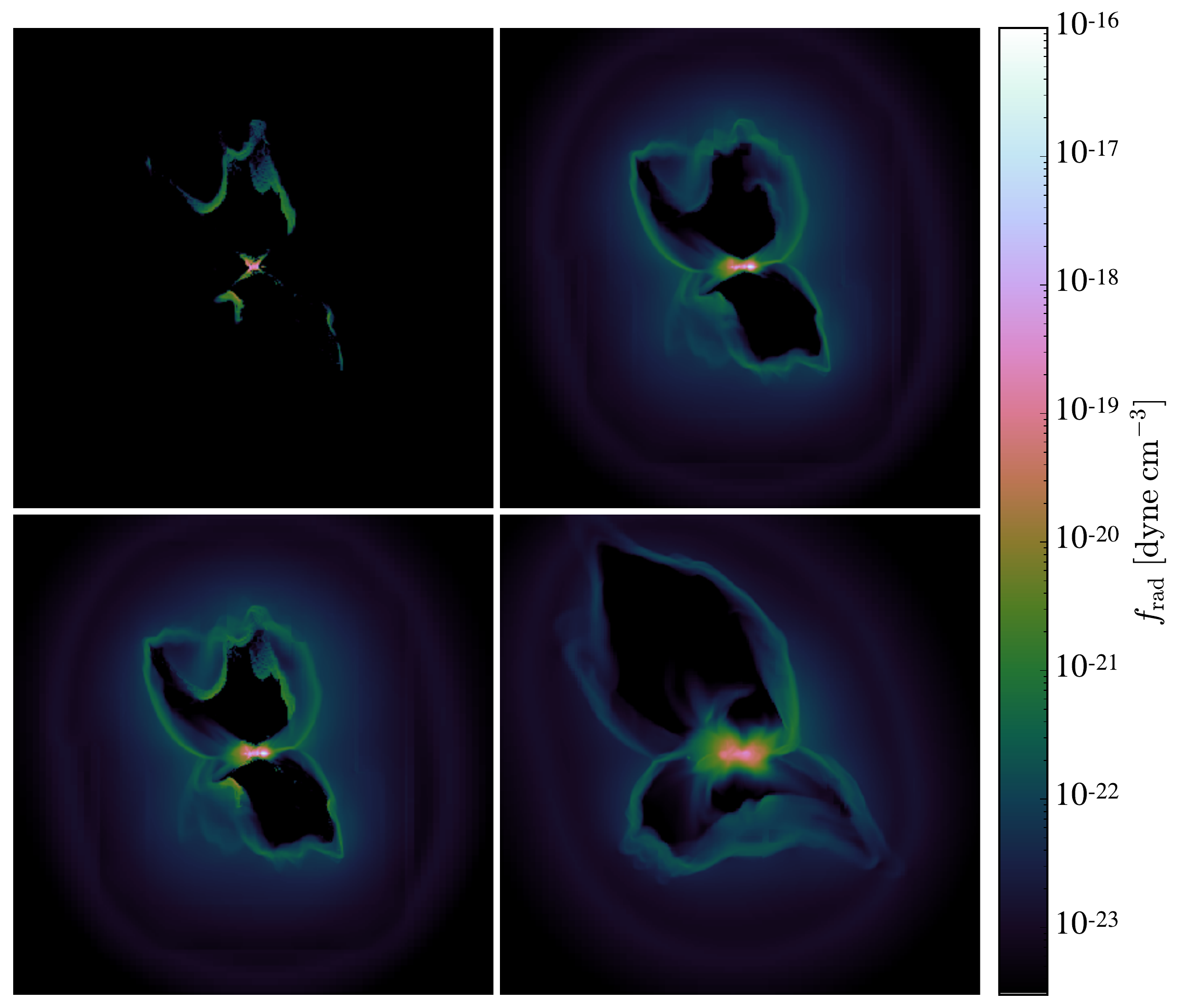}}
\caption{
\label{fig:fradcomp}
Volume-weighted projection plots of the radiation force densities along the $yz$-plane for the direct radiation field (top left panel) and diffuse radiation field (top right panel) in run \noturb\ and total radiation field in runs \noturb\ (lower left panel) and \nort\ (lower right panel), respectively, when the star has a mass of 36.1 $M_{\rm \odot}$. Each projection covers a depth of 500 AU and area of (12,000 AU)$^2$. The center of each panel corresponds to the location of the most massive star.
}
\end{figure}

\begin{figure*}
\centerline{\includegraphics[trim=0.2cm 0.2cm 0.2cm 0.2cm,clip,width=0.7\textwidth]{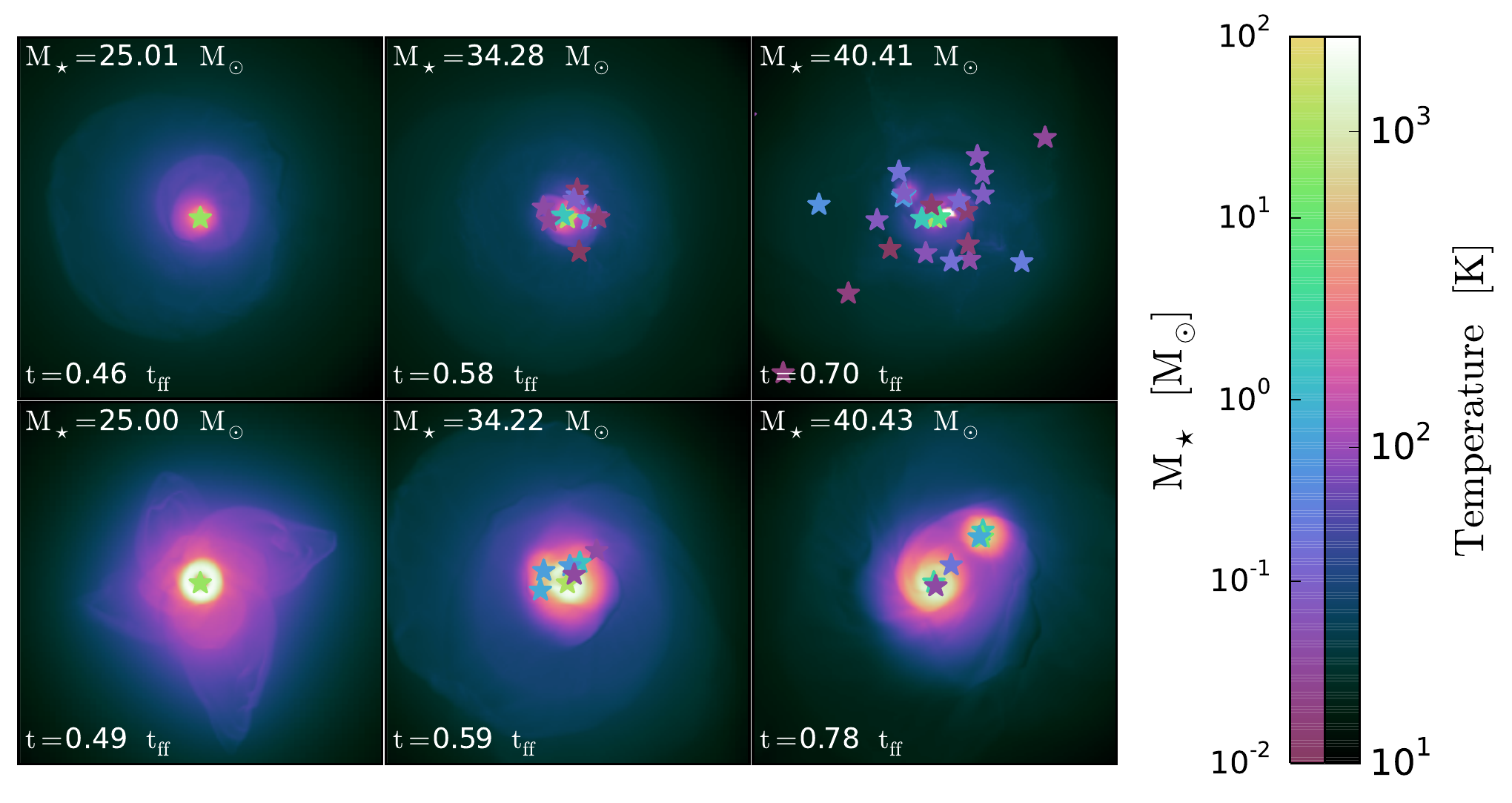}}
\caption{
\label{fig:tempdisk}
Density-weighted mean projected temperature for the accretion disks formed in runs \noturb\ (top row) and \nort\ (bottom row), respectively. Each panel is a projection that is (5000 AU)$^2$ in size and is projected over a height of 1000 AU above and below the massive star. The most massive star is at the center of each panel, and stars with masses greater than 0.01 $M_{\rm \odot}$ are over-plotted. 
}
\end{figure*}

\subsection{Collapse of Turbulent Pre-stellar Cores}
\label{sec:turb}

Next we describe our results for run \turb, which follows the same initial conditions as run \noturb\ except that the core is not initially placed in solid-body rotation and is instead seeded with a weakly turbulent velocity profile with a 1D velocity dispersion of $\sigma_{\rm 1D}=0.4 \; \rm km \, s^{-1}$. At the end of this simulation the most massive star has a mass of $61.63\; M_{\odot}$. We ran the simulation for a time of $t=0.87 \; t_{\rm ff}$.


\subsubsection{Evolution of Radiation Pressure Dominated Bubbles}
\label{sec:noturbbubble}
\begin{figure*}
\centerline{\includegraphics[trim=0.2cm 0.2cm 0.2cm 0.2cm,clip,width=0.7\textwidth]{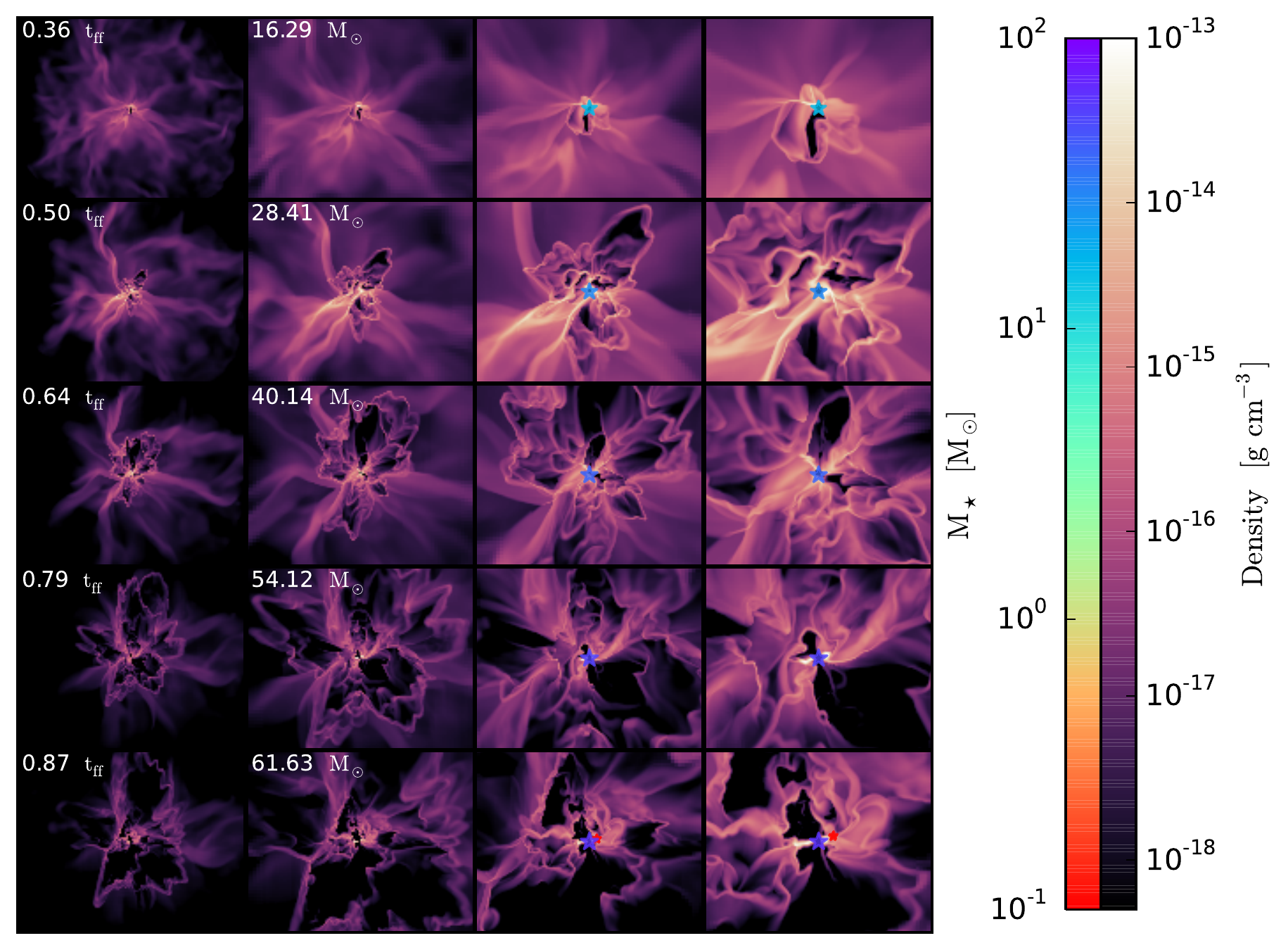}}
\caption{
\label{fig:turbx}
Same as Figure \ref{fig:noturbx}, but for run \turb. The center of each panel corresponds to the position of the most massive star.
}
\end{figure*}

Figure \ref{fig:turbx} shows density slices along the $yz$-plane for a series of snapshots from run \turb. The initial turbulence leads to a clumpy and filamentary core density structure that begins to collapse and forms a star. As the core continues to collapse the star is primarily fed by dense filaments and clumpy material. We first see a radiation pressure dominated bubble begin to expand when the star is $\sim10 \; M_{\rm \odot}$ but it is quickly quenched by the dense, inflowing material. Furthermore, these bubbles instantly go RT unstable and deliver material to the star (i.e, within the 80 AU accretion radius of the sink particle). This can be seen within the radiation pressure dominated bubble interiors shown in Figure \ref{fig:turbx} because the size scale of the density fluctuations within the bubbles is smaller than the density perturbations in the initial turbulence surrounding the bubbles, thus suggesting that the interior bubble material has become RT unstable. We find that the growth of a radiation driven bubble is continuously suppressed by the flux of the infalling filamentary and RT unstable material until the star reaches a mass of $\sim21.7 \; M_{\rm \odot}$ at time $t=0.42 \; t_{\rm ff}$. At this time the direct radiation pressure from the accreting star is able to effectively push material away from the star.  However, material is not fully evacuated along the polar directions of the star.

\begin{figure}
\centerline{\includegraphics[trim=0.2cm 0.2cm 0.2cm 0.2cm,clip,width=0.85\columnwidth]{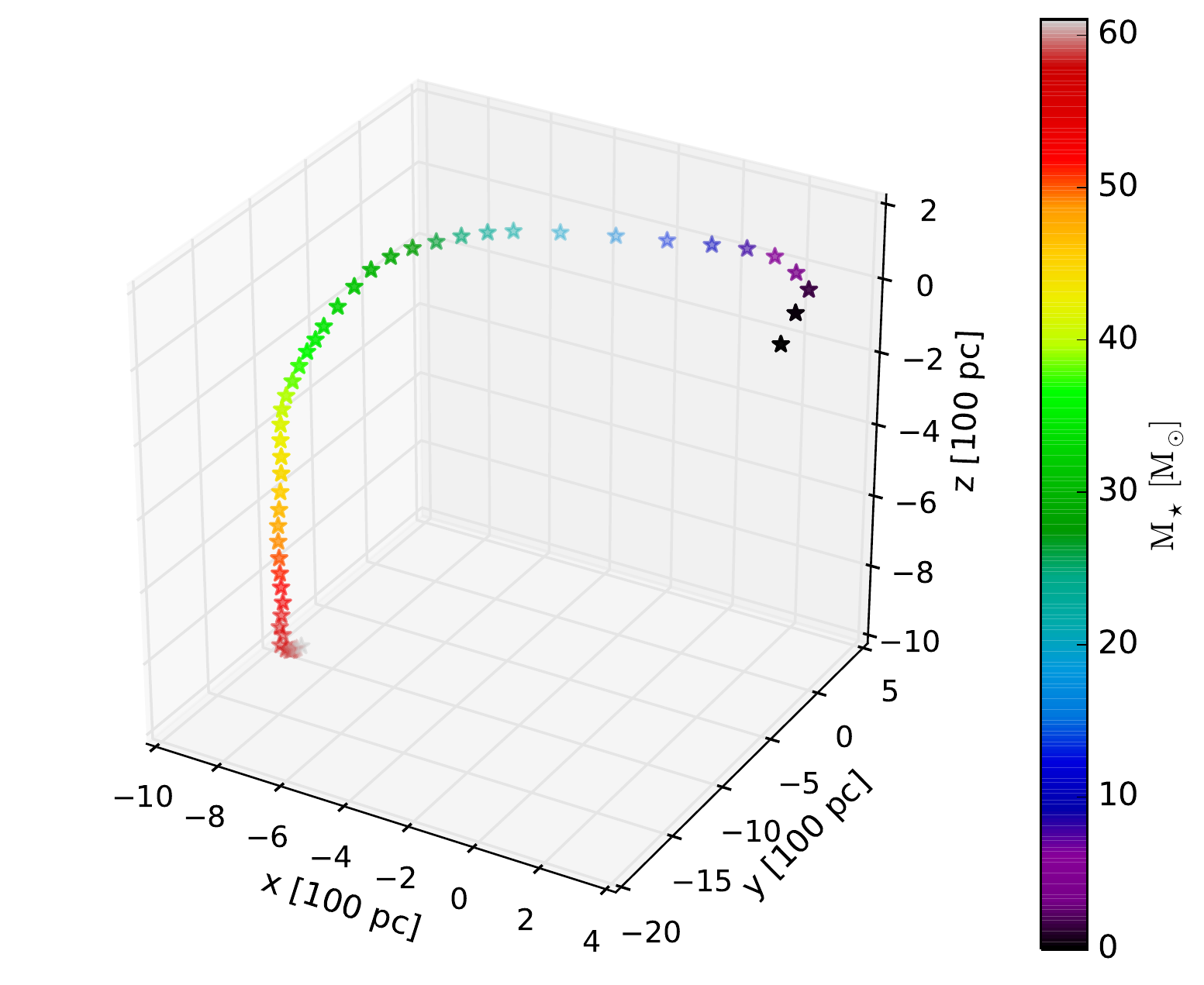}}
\caption{
\label{fig:starpos}
Three-dimensional position of primary star in run \turb.
}
\end{figure}

\begin{figure*}
\centerline{\includegraphics[trim=0.2cm 0.2cm 0.2cm 0.2cm,clip,width=0.8\textwidth]{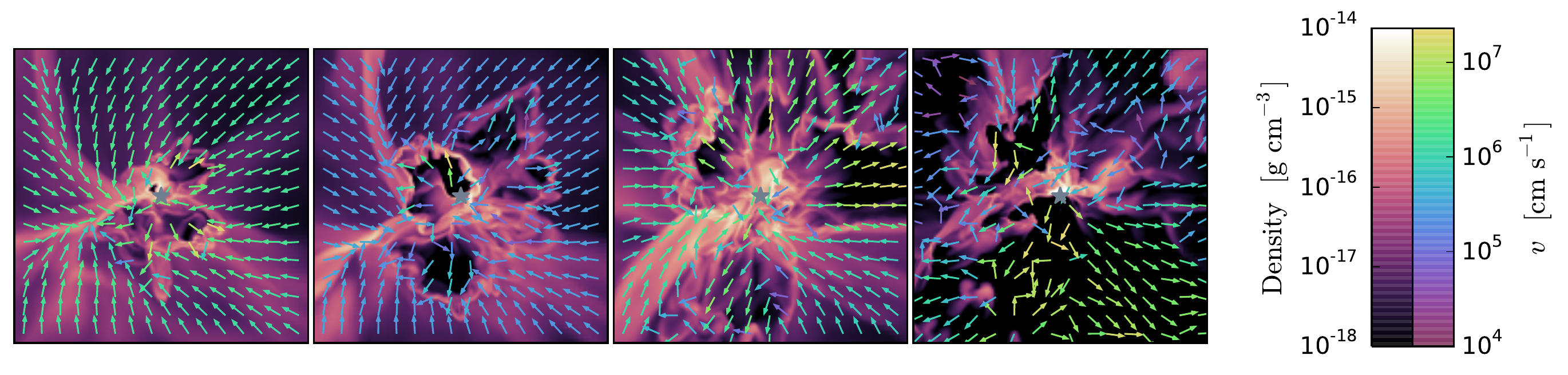}}
\caption{
\label{fig:turbvel}
Density slices along the $yz$-plane with velocity vectors over plotted for run \turb\ when the massive star is (from left to right) 23.82 $M_{\rm \odot}$ at $t=0.43\, t_{\rm ff}$, 30.03 $M_{\rm \odot}$ at $t=0.53\, t_{\rm ff}$, 41.08 $M_{\rm \odot}$ at $t=0.65\, t_{\rm ff}$, and 61.63 $M_{\rm \odot}$ at $t=0.87\, t_{\rm ff}$, respectively. The region plotted is (10,000 AU)$^{2}$ with the most massive star (over plotted gray star) located at the center of each panel.
}
\end{figure*}

This quick onset of RT instabilities can be attributed to the initial turbulence. The turbulent gas seeds the growth of these instabilities. In addition, when the star is below $\sim30 \; M_{\rm \odot}$ the star is moving rapidly in the core because the accreting material carries momentum. Figure \ref{fig:starpos} shows that the stellar velocity decreases as the stellar mass increases. The overall accretion flow onto the star is not spherically symmetric and thus the asymmetrical momentum deposition to the star by the accreting gas causes the star to move a significant distance in the collapsing core, a property also observed in the massive star formation simulations presented in \citet{Cunningham2011a}. Throughout the simulation the star travels a distance of 1968 AU from its initial position. The combination of the movement of the star, dense filaments accreting onto the star, and RT instabilities delivering material to the star limit the growth and stability of radiation pressure dominated bubbles around the star. When the star has a mass greater than $\sim 30 \; M_{\rm \odot}$ radiation pressure begins to evacuate low-density material away from the star but dense filaments and material that become RT unstable continue to fall onto the star. This effect is demonstrated in Figure \ref{fig:turbvel} that shows density slices along the $yz$-plane for a series of snapshots from run \turb\ with velocity vectors over-plotted. 

Low density bubbles do not begin to appear until the star has reached a mass of $\sim 35 \; M_{\rm \odot}$ at time $t=0.59 \; t_{\rm ff}$ and these bubbles are larger than those in run \noturb\ at a similar mass ($\sim$4000 AU versus $\sim$3000 AU in run \noturb). At this mass the stellar luminosity is large enough to push away the infalling material more effectively. The bubbles then become episodic, expanding and collapsing as the ram pressure of the inflow rises and falls due to the turbulence. This behavior continues throughout the rest of the simulation but the bubbles survive longer and expand as the stellar luminosity increases. At the end of run \turb, when the star has a mass of $61.63\; M_{\odot}$ at time $t=0.87 \; t_{\rm ff}$, most of core has been evacuated by radiation pressure along the polar directions of the star but material is still being fed to the star along directions that are perpendicular to the poles of the star (e.g., see bottom panel of Figure \ref{fig:turbx}).

\begin{figure}
\centerline{\includegraphics[trim=0.2cm 0.2cm 0.2cm 0.2cm,clip,width=0.85\columnwidth]{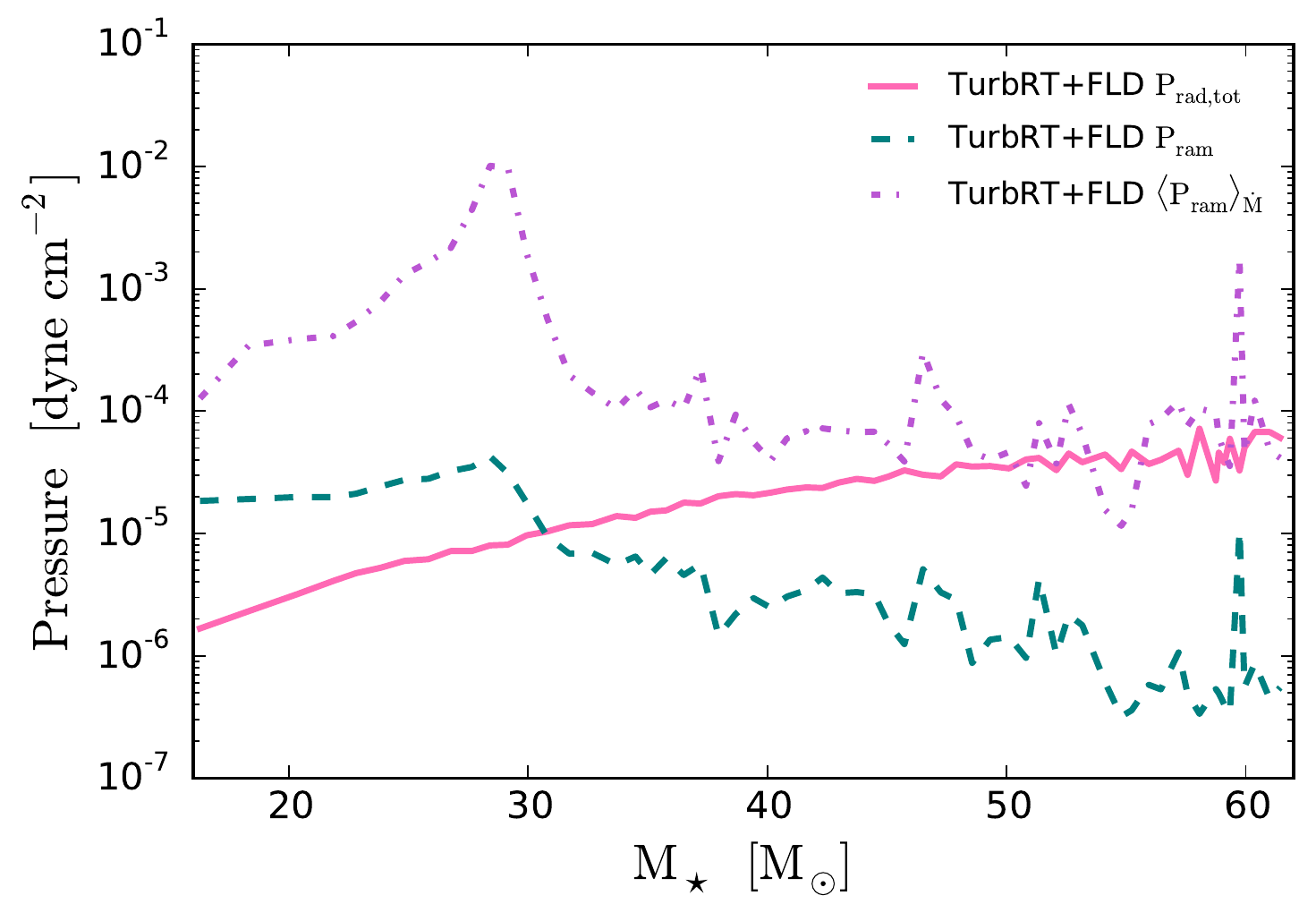}}
\caption{
\label{fig:pres}
Comparison of the direct radiation pressure including contributions from the stellar and accretion radiation fields (pink solid line) and the area-weighted and mass-weighted ram pressure (teal dashed and purple dot-dashed lines, respectively) from inflowing material taken over a 1000 AU sphere surrounding the accreting primary star for run \turb. See main text for full details on how these averages are defined.
}
\end{figure}

To understand this behavior quantitatively, it is helpful to compare the pressure of the stellar radiation field to the ram pressure of the inflow. Consider a sphere 1000 AU in radius centered on the most massive star. To understand the force balance in the problem, we compute three mean pressures on this sphere: the direct radiation pressure (including the accretion and stellar luminosities, averaged over area), the area-weighted mean ram pressure, and the mass-flux weighted mean ram pressure for inflowing material. The former quantity is defined as
\begin{equation}
P_{\rm rad} = \frac{L_{\rm \star}+L_{\rm acc}}{4 \pi r^2 c}
\end{equation}
and the latter two quantities are defined by
\begin{equation}
\langle P_{\rm ram}\rangle = \frac{\int \rho v_r^2 w \, dA}{\int w \, dA},
\end{equation}
where $v_r$ is the radial velocity, and the weighting function $w$ is unity for the area-weighted average, and is $\rho v_r$ for the mass flux-weighted average where we only include contributions from inflowing material. We include the mass-flux weighted mean ram pressure because it is a better representation of the ram pressure of the material that can be accreted onto the star. We plot these three quantities as a function of time in Figure \ref{fig:pres}. We see that the radiation pressure overwhelms the area-averaged ram pressure by the time the star reaches $\sim 30$ $M_\odot$, but that the mass flux-weighted mean ram pressure is roughly an order of magnitude higher. Thus, even though the radiation pressure is stronger than gas pressure when averaged over $4\pi$ sr, turbulence causes the mass flow onto the star to be concentrated in narrow filaments that have much greater ram pressure, and are much harder to stop. It is not until the star reaches $\sim 50$ $M_\odot$ that its radiation pressure becomes comparable to the mass flux-weighted ram pressure, and even then there are still periods when the ram pressure rises and is sufficient to punch through the radiation and deliver mass. This episodic rise in the ram pressure (which is mirrored in the density field by the episodic collapse of the radiation-dominated bubbles) is a direct result of RT instability, accelerated and seeded by the pre-existing turbulence.

\subsubsection{Accretion Disk Evolution}
\begin{figure*}
\centerline{\includegraphics[trim=0.2cm 0.2cm 0.2cm 0.2cm,clip,width=0.7\textwidth]{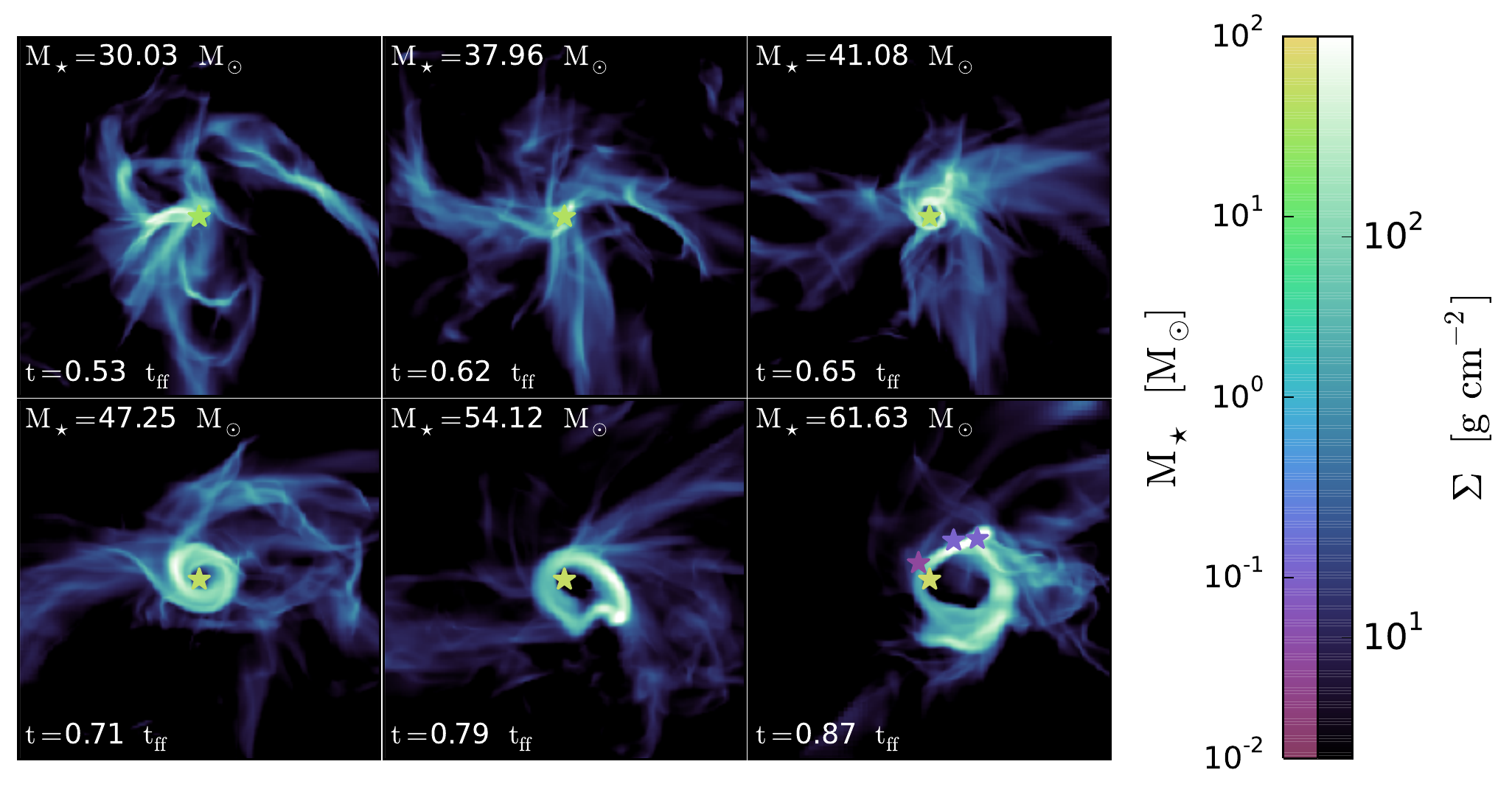}}
\caption{
\label{fig:turbz}
Same as Figure \ref{fig:noturbz}, but for run \turb.
}
\end{figure*}

\begin{figure}
\centerline{\includegraphics[trim=0.2cm 0.2cm 0.2cm 0.2cm,clip,width=0.85\columnwidth]{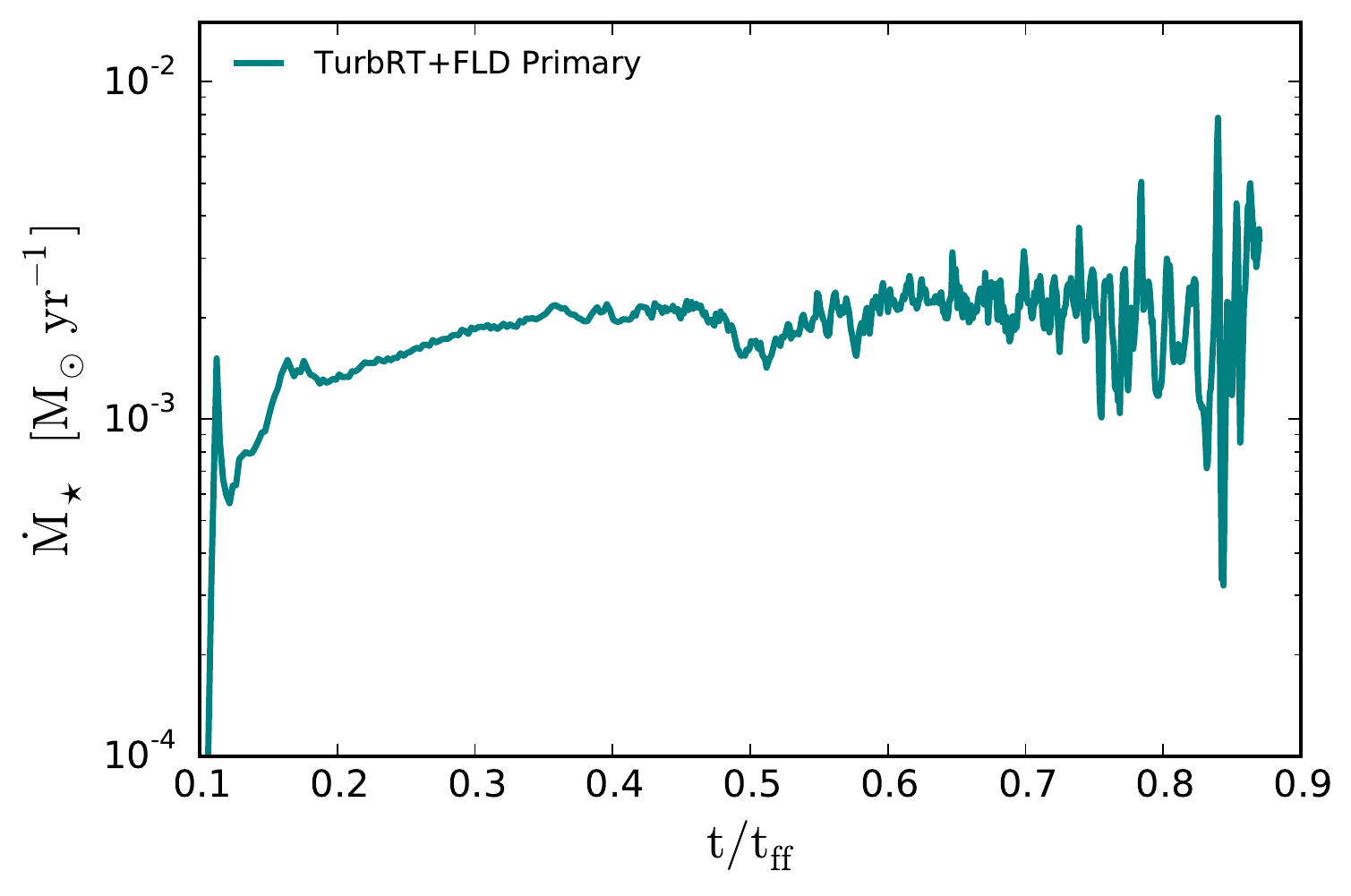}}
\caption{
\label{fig:turbmdot}
Primary star accretion rate as a function of time for run \turb.
}
\end{figure}

Figure \ref{fig:turbz} shows the growth and evolution of the accretion disk that forms around the massive star in run \turb. Our results show that a thick accretion disk begins to form around the massive star when it has reached a mass of  $\sim 41 \; M_{\rm \odot}$ at time $t=0.65 \, t_{\rm ff}$ (i.e., an accretion disk with a radius larger than the 80 AU accretion zone radius of the sink particle). Up until this point material is primarily fed to the star through dense filaments and RT instabilities whose infall is not suppressed by radiation pressure. Figure \ref{fig:turbmdot} shows the accretion rate onto the primary star as a function of time. The accretion rate is relatively smooth up until a time of $t=0.5 \; t_{\rm ff}$. After this time, when the star has a mass of $\sim $28 $M_{\rm \odot}$, RT instabilities and dense filaments supply most of the mass onto the star leading to a chaotic accretion flow onto the star. However, when a thick accretion disk forms, at $t \approx 0.65 \; t_{\rm ff}$, the accretion rate onto the star becomes much more chaotic because gravitational instabilities in the disk funnel gas to the star while dense filaments and RT unstable material are delivered to the disk. The disk soon becomes unstable and begins to fragment when the primary star has a mass of $57.62\; M_{\odot}$ at time $t=0.83 \; t_{\rm ff}$. At the end of run \turb\ the primary star has three low-mass companion stars with masses between 0.034-0.11 $M_{\rm \odot}$. Again we form a hierarchical system consisting of a massive primary and a series of much less massive mass companions similar to run \noturb.

\section{Discussion}
\label{sec:discussion}
The purpose of this work is to understand how mass is delivered to massive stars during their formation. Primarily,  we are interested in determining if the radiation pressure dominated bubbles that expand away from the star become RT unstable and if these instabilities contribute to disk accretion or direct accretion onto the star. To answer this question, and compare our work with previous 3D simulations of the formation of massive stars, we developed a new highly parallel hybrid radiation algorithm that models the direct radiation pressure from stars with a multi-frequency long-characteristics ray tracing solve coupled to (gray) flux-limited diffusion to model the re-emission and processing by interstellar dust in the \orion\ radiation- simulation code \citep{Rosen2016a}. With this new tool we have performed the collapse of initially laminar and turbulent massive star-forming cores. 

Our results lead to two key differences from the simulations presented in \citet{Kuiper2011a, Kuiper2012a} and \citet{Klassen2016a}. The first crucial difference, which we discuss in Section \ref{sec:rrt}, is that we find that the radiation pressure dominated bubbles that expand around the accreting massive star become unstable and deliver mass to the star-disk system for both initially laminar and turbulent cores. The second difference we find, which we address in Section \ref{sec:disk}, is that inclusion of direct radiation pressure leads to unstable accretion disks that fragment into a hierarchical system consisting of a massive primary and a series of much less massive companions. Finally, we also find if the pre-stellar core is initially turbulent the growth of radiation pressure dominated bubbles are suppressed at early times as compared to massive stars that form out of initially laminar cores. For initially turbulent cores, we find that most of the mass is supplied to the star via dense filaments and RT instabilities rather than extended disk accretion. We discuss these differences in Section \ref{sec:flash}.

\subsection{Revisiting Rayleigh Taylor Instabilities}
\label{sec:rrt}
\citet{Kuiper2011a, Kuiper2012a} and \citet{Klassen2016a} find that the expanding radiation pressure dominated bubbles that form around accreting massive stars are stable and the massive star is only fed by disk accretion. They suggest that the bubbles that form in the simulation presented in \citet{Krumholz2009a} develop RT instabilities because only the diffuse dust-reprocessed radiation field is modeled and therefore the true radiation pressure is underestimated. These authors conclude that inclusion of the direct radiation field from the star leads to a larger radiation pressure resulting in stable bubbles that are not subject to collapse. In contrast, we find that, while an improved treatment of the direct radiation field does lead to a larger radiation pressure it only \textit{delays} the onset of substantial RT instabilities that are capable of delivering mass to the star-disk system in the case of a laminar core, it does not prevent them entirely. At late times these instabilities non-linearly in regions that are shielded from the direct radiation field and deposit material to the star-disk system. This material can then be fed to the accreting star. We would like to understand the origin of this difference in results, though we caution that, in the light of our results for run \turb, this discussion is somewhat academic. This run shows that, in a realistic, turbulent core, the flow is ``born" RT unstable, in the sense illustrated by Figure \ref{fig:pres} -- a configuration whereby the angle-averaged radiation force is stronger than gravity and ram pressure, but the majority of the mass flux arrives over a small solid angle where the ram pressure force is stronger than the radiation force.

Since \citet{Kuiper2011a, Kuiper2012a} and \citet{Klassen2016a} hypothesize that the absence of RT instability in their simulations was due to the direct radiation force, we begin this investigation by examining when the presence of direct radiation pressure could possibly make a difference to the outcome. We therefore investigate when the gravitational force per unit area,  $f_{\rm grav}(r)=GM_{\rm \star} \Sigma/r^2$, exceeds the direct radiative force per unit area, $f_{\rm rad,dir}(r)=L_{\rm \star}/(4 \pi r^2 c)$. Here $M_{\rm \star}$ and $L_{\rm \star}$ are the mass and luminosity of the massive star, respectively; and $\Sigma(r)=\int^r_0 \rho(r') dr'$ is the column density of a slab of core material as seen by the star. The relative importance of direct radiation force and gravity can be described in terms of the Eddington ratio given in Section \ref{sec:intro} where we now ignore the contribution of the trapped diffuse radiation force, $f_{\rm edd,dir} = f_{\rm rad,dir}/f_{\rm grav}$, given by
\begin{equation}
\label{eqn:fedd}
f_{\rm edd,dir} = 7.7 \times 10^{-5} \left( \frac{L_{\rm \star}}{M_{\rm \star}} \right)_{\rm \odot} \left( \frac{\Sigma}{1 \; \rm{g \; cm^{-2}}} \right)^{-1}. 
\end{equation}
The notation $(...)_{\rm \odot}$ denotes that $L_{\rm \star}$  and $M_{\rm \star}$ are in units of $L_{\rm \odot}$ and $M_{\rm \odot}$, respectively. A value of $f_{\rm edd,dir} \gtrsim 1$ implies that direct radiation pressure is dynamically dominant, while a value of $f_{\rm edd} < 1$ implies that gravity dominates. In the latter regime the force exerted by the \textit{diffuse} radiation field may still exceed the force of gravity, but in such regions we expect RT instability to occur \citep{Jacquet2011a}. Locations where $f_{\rm edd,dir} < 1$ can therefore collapse and deliver mass to the star-disk system. 

\begin{figure}
\centerline{\includegraphics[trim=0.2cm 0.2cm 0.2cm 0.2cm,clip,width=0.95\columnwidth]{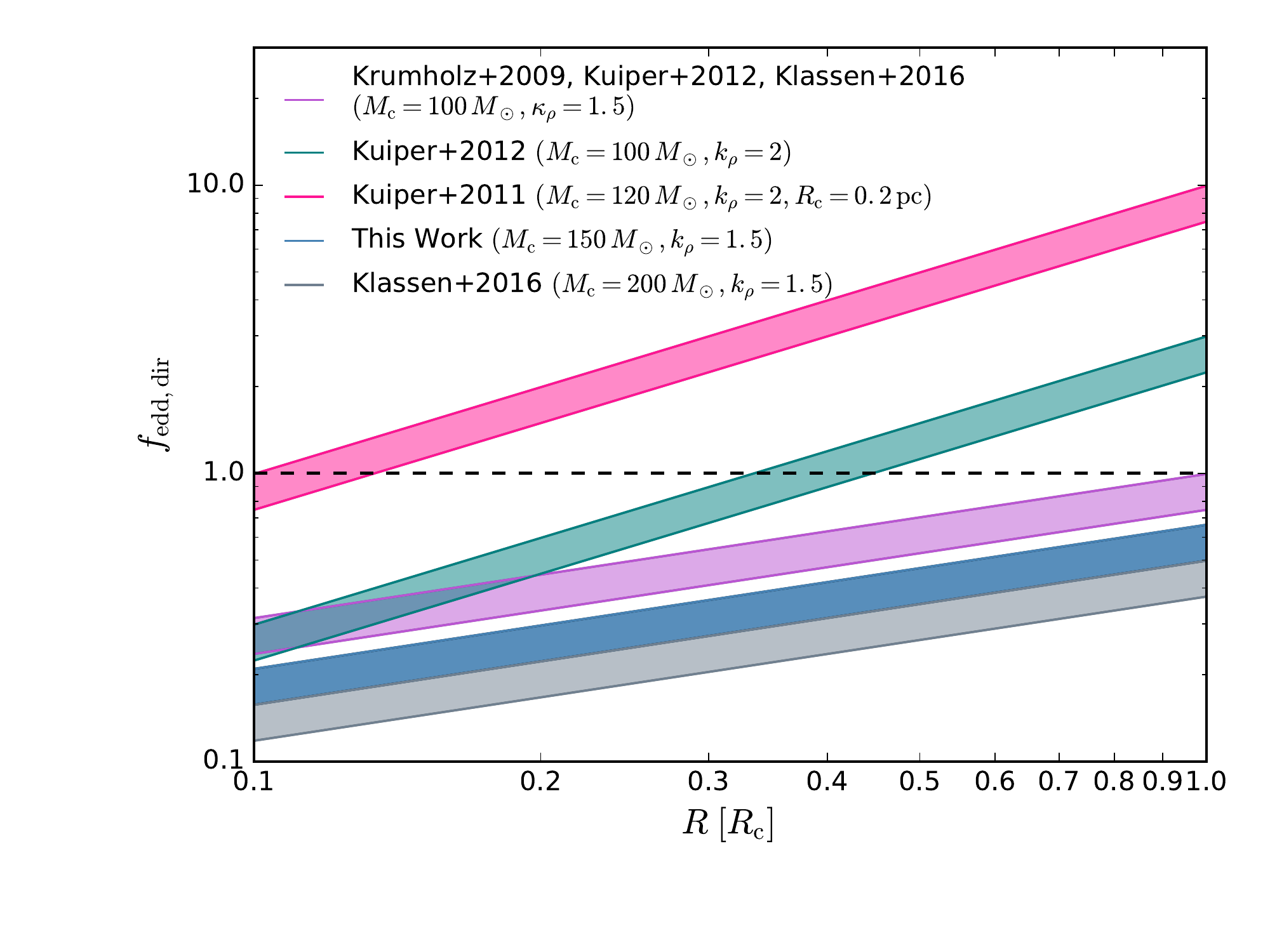}}
\caption{
\label{fig:feddcomp}
Eddington ratio associated with the direct radiation field ($f_{\rm edd,dir}$) for the initial core properties listed in Table \ref{tab:comp} as a function of radius within the core. Shaded regions denote $L_{\rm \star}/M_{\rm \star}$ values computed for a zero age main sequence star between 35 (bottom line) and 45 (top line) $M_{\rm \odot}$ in mass. The horizontal black dashed line denotes where $f_{\rm edd,dir}=1$.
}
\end{figure}

We compute $f_{\rm edd,dir}$ as a function of position within the initial pre-stellar cores modeled by \citet{Krumholz2009a}, \citet{Kuiper2011a}, \citet{Kuiper2012a}, \citet{Klassen2016a}, and this work (using the core properties listed in Table \ref{tab:comp}) for light to mass ratios appropriate for zero age main sequence stars with masses $M=35 - 45$ $M_\odot$, and plot the results in Figure \ref{fig:feddcomp}. For the purposes of this computation, note that the core density profile in all of these simulations is $\rho(R) = \rho_{\rm edge} (R/R_c)^{-k_{\rm \rho}}$ where $\rho_{\rm edge} = (3-k_\rho)M_{\rm c}/(4\pi R^3_{\rm c})$ for a pre-stellar core with mass $M_{\rm c}$ and radius $R_{\rm c}$. The resulting column density, as seen by the star at a distance $R$, is $\Sigma(R)=\rho_0 R_c^{k_{\rm \rho}} R^{1-k_{\rm \rho}}/(k_{\rm \rho}-1)$. Of these simulations only those presented in \citet{Krumholz2009a} and this work find that the radiation-pressure dominated bubble shells become RT unstable, and Figure \ref{fig:feddcomp} makes it clear that at least part of this discrepancy is simply a matter of initial conditions. We find that the cores with $k_{\rho}=2$ presented in \citet{Kuiper2011a} and \citet{Kuiper2012a} have $f_{\rm edd,dir} > 1$ over a substantial portion of their radial extent, as a result of the cores' steeper density profile, and lower overall surface density. As a result, direct radiation pressure alone, without any assistance from the diffuse reprocessed radiation field, is sufficient to expel material from the cores simulated by \citet{Kuiper2011a}, and possibly also \citet{Kuiper2012a}. It is not surprising, given such a setup, that RT instability does not develop -- the diffuse radiation field never matters, because direct radiation alone guarantees a net positive radial acceleration.

While this simple argument explains why the cores simulated by \citet{Kuiper2011a} and perhaps \citet{Kuiper2012a} never develop RT instabilities, it does not explain the discrepancy between our results and those of \citet{Klassen2016a}, who have direct Eddington ratios comparable to those in our simulation. One key difference between the work of those authors and the simulations we present here is refinement of the bubble shell. Refining the bubble shell is crucial for studying the growth of RT instabilities in massive star formation simulations because the amplitude $\eta$ of linear perturbations (which is the relevant regime for the non-turbulent simulations) grows with time as $\eta(t) \propto \exp{\left(\omega t \right)}$, where $\omega \propto \lambda^{-1/2}$ and $\lambda$ is the wavelength of the perturbation \citep{Jacquet2011a}. Thus for radiation RT instability, as for classical hydrodynamic RT instability, the smallest perturbations grow fastest. However, perturbations can only grow if they are resolved. 

\begin{figure}
\centerline{\includegraphics[trim=0.2cm 0.2cm 0.2cm 0.2cm,clip,width=0.95\columnwidth]{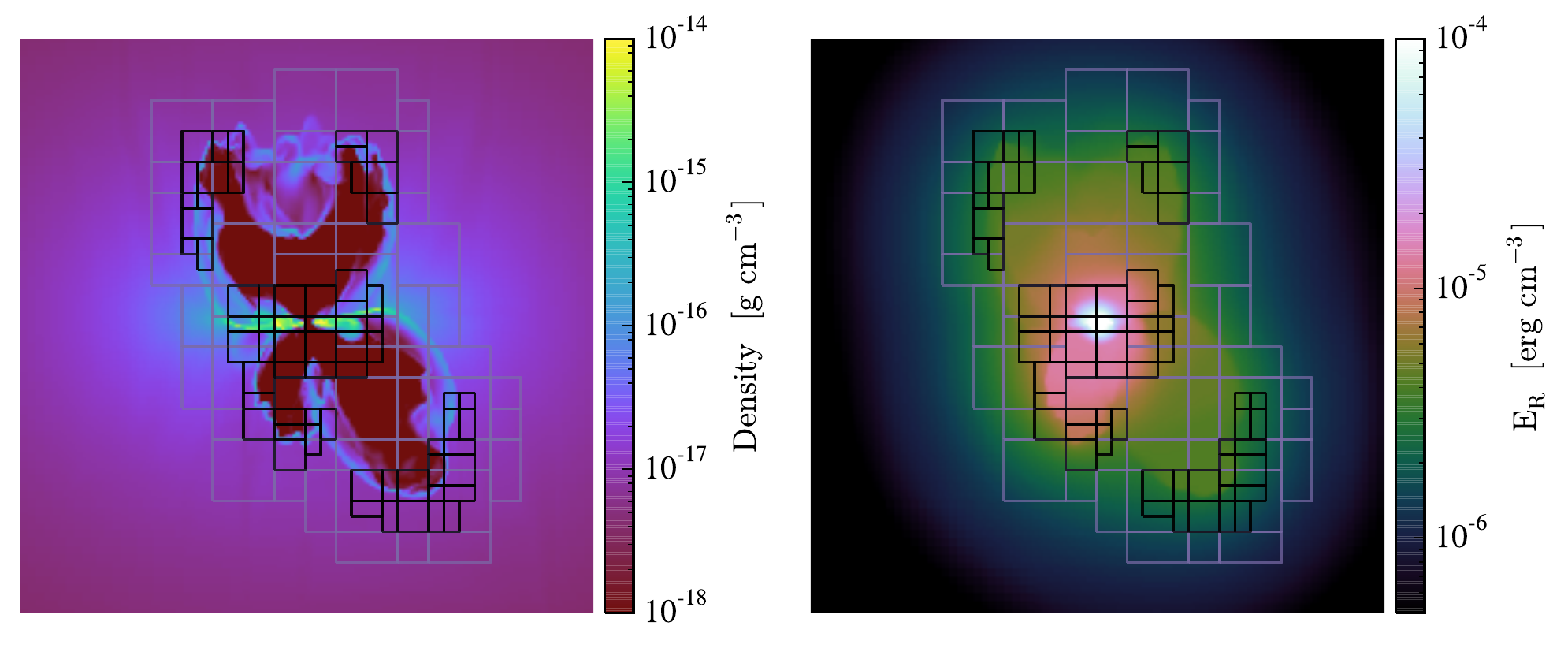}}
\caption{
\label{fig:grids}
Density and radiation energy density slices along the $yz$-plane when the star has a mass of 40.4 $M_{\rm \odot}$ at $t=0.7 \; t_{\rm ff}$. Level 4 grids (40 AU resolution -- gray rectangles) and level 5 grids (20 AU resolution -- black rectangles) are over-plotted. The region plotted is (12,000 AU)$^{2}$ and the center of each panel corresponds to the primary star location.
}
\end{figure}

Based on this analysis, the lack of RT instability in the work of \citet{Kuiper2012a} and \citet{Klassen2016a}, and its presence in our simulations, is likely a resolution effect. We can make this point quantitatively as follows. In this work, and the work of \citet{Krumholz2009a}, we adaptively refine on the Jeans length and locations where the gradient of the radiation energy density exceeds 15\%. Figure \ref{fig:grids} shows density and radiation energy density slices, with the level 4 and 5 grids over-plotted, at the time when the star in run \noturb\ has a mass of 40.4 $M_{\rm \odot}$ at time 0.7 $t_{\rm ff}$. This Figure shows that the radiation pressure dominated bubbles are refined up to level 4 (40 AU resolution) and that most of the bubble shells have level 5 refinement (20 AU resolution). The level 5 refinement for the bubble shell is due to our radiation energy density gradient refinement criterion, because this refinement condition is triggered by the sharp gradient in the radiation energy density at the shell location (i.e., where it transitions from optically thin to optically thick material). 

In contrast, \citet{Kuiper2012a} use a non-adaptive spherical grid that provides much higher resolution than we achieve near the star, but that coarsens rapidly at large distances. The grid has a resolution of $5.625^\circ$ in the $\theta$ direction, which at a distance of 4,000 AU, roughly the sizes of our bubbles at the point where they become unstable, corresponds to $\approx 400$ AU. Thus their resolution is a factor of $\sim 20$ lower than ours, and the linear growth rate is a factor of $\approx 4.5$ lower. This may push the development of the instability back to times longer than the time required for all the mass to be accreted. The situation for \citet{Klassen2016a} is similar. While they do have adaptivity, they refine only on Jeans length and not on gradients in the radiation energy density, and thus their bubble walls are at much lower resolution than the peak they achieve. Visual inspection of their published results (their Figure 12) suggests that a typical resolution in their bubble walls is 160 AU, giving a growth time $\approx 3$ times longer than we have.

\begin{table*}
	\begin{center}
		\caption{
	\label{tab:comp}
Comparison of the initial laminar pre-stellar core conditions from this work and previous numerical work.
}	
	\begin{tabular}{ l c c c c c c}
	\hline	
	Work &  $M_{\rm c} \; [\rm{M_{\rm \odot}}]$ & $R_{\rm c} \; \rm{[pc]}$  & $k_{\rm \rho}$ & $\Sigma_{\rm c}$\tablenotemark{a} [g cm$^{-2}$] &  RTI?\tablenotemark{b}
	 \\
	\hline
	\hline
	This work & 150 & 0.1 & 1.5  & 1 & Yes \\
	\citet{Klassen2016a} & 100, 200 & 0.1 & 1.5 & 0.67, 1.33 & No\\
	\citet{Kuiper2012a}  & 100 & 0.1 & 1.5, 2 & 0.67  & No\\
	\citet{Kuiper2011a}  & 120 & 0.2 & 2 &  0.2  & No\\
	\citet{Krumholz2009a} & 100 & 0.1 & 1.5 & 0.67 & Yes\\
	\hline
	\end{tabular}
	\end{center}
	\tablenotetext{a}{$\Sigma_c \equiv M_c / \pi R_c^2$}
	\tablenotetext{b}{Rayleigh Taylor instabilities present?}
	\tablerefs{\citet{Krumholz2009a,Kuiper2011a,Kuiper2012a, Klassen2016a}}
\end{table*}

\begin{figure*}
\centerline{\includegraphics[trim=0.2cm 0.2cm 0.2cm 0.2cm,clip,width=0.8\textwidth]{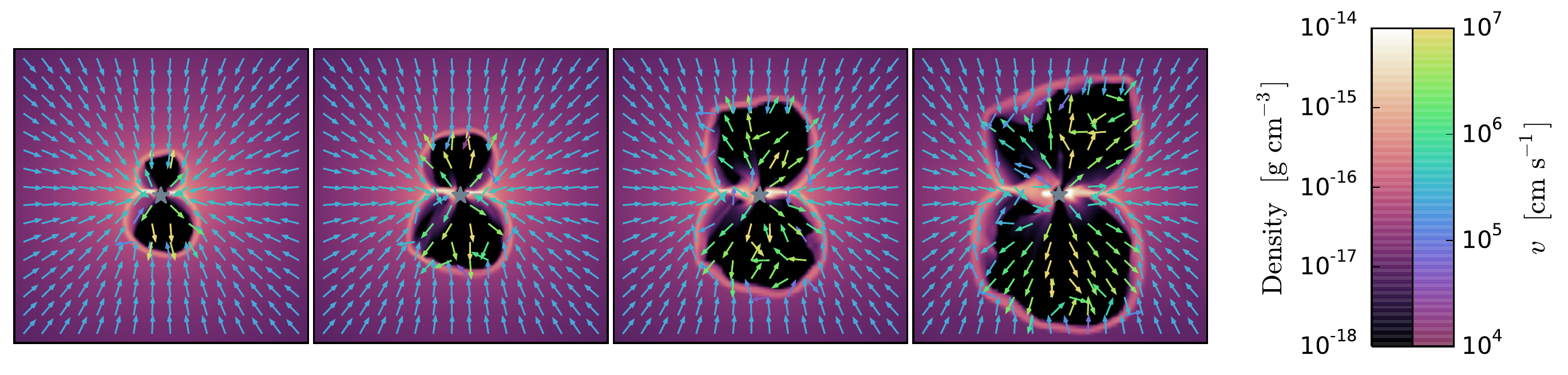}}
\caption{
\label{fig:noturblr}
Density slices along the $yz$-plane with velocity vectors over-plotted for run \noturblr\ when the massive star is (from left to right) 24.70 $M_{\rm \odot}$ at $t=0.43\, t_{\rm ff}$, 30.05 $M_{\rm \odot}$ at $t=0.46\, t_{\rm ff}$, 34.22 $M_{\rm \odot}$ at $t=0.50\, t_{\rm ff}$, and 40.34 $M_{\rm \odot}$ at $t=0.57\, t_{\rm ff}$, respectively. The region plotted is (8,000 AU)$^{2}$ with the most massive star (over plotted gray star) located at the center of each panel.
}
\end{figure*}

We can also test the resolution hypothesis directly. To do so, we perform a low resolution run, run \noturblr, which has four levels of refinement (40 AU maximum resolution) and only refines on the Jeans length. Thus run \noturblr\ uses the same refinement criteria as \citet{Klassen2016a}. This reduction in refinement criteria ensures that the shell will be poorly resolved. Figure \ref{fig:noturblr} shows density slices along the $yz$-plane showing the evolution of the expanding radiation dominated pressure bubbles for run \noturblr. Comparison of Figure \ref{fig:noturblr} with Figure \ref{fig:noturbx} demonstrates that when the shell is poorly resolved RT instabilities are unlikely to develop.  In run \noturb\ we found that RT instabilities began to have noticable growth when the star was $\sim30 \; M_{\rm \odot}$ whereas the bubbles in run \noturblr\ remain stable until the star has reached a mass of $\sim37.5 \; M_{\rm \odot}$, and even after this point the instability is clearly less violent and delivers less mass than in the higher resolution run. Clearly resolution matters a great deal for the development of RT instability.

The late-onset instability in our low resolution run also points to one more potentially important difference between our simulations and those of \citet{Klassen2016a}. The instability in run \noturblr\ first appears when the left side of the disk becomes flared and most of the direct radiation is absorbed by the disk, shadowing the left side of the bubble. Shadowing of the direct radiation field is clearly an important process. Our direct radiation treatment uses the method of long characteristics that is more accurate than the method of hybrid characteristics used in \citet{Klassen2016a}. Far away from the source, as the rays cross many grids, this method will not resolve sharp shadows as well as long characteristics, and will likely underestimate the asymmetry in the direct radiation field thus suppressing the development of RT instabilities.

\subsection{Revisiting Disk Fragmentation}
\label{sec:disk}
Most massive stars are found in multiple systems. \citet{Chini2012a} performed a spectroscopic study of massive stars and found that  $> 82\%$ of stars with masses greater than 16 $M_{\rm \odot}$ belong to close binary systems. Likewise, \citet{Sana2014a} found that all massive main-sequence stars in their sample are in tight binaries or higher order multiples. Both authors conclude that this large binary fraction originates from the formation process rather than direct capture. In agreement with these observations, we formed multiple systems in runs \noturb\ and \turb\ where companions form from disk fragmentation. At the end of each run we are left with a hierarchical system consisting of a massive primary and a series of low-mass companions. Contrary to our results, \citet{Klassen2016a} only form a single massive star in each of their simulations while \citet{Krumholz2009a} form a massive binary system. We attribute these differences in stellar multiplicity to be due to the different sink particle creation and merging algorithms employed.

The accretion disks formed in the work of \citet{Klassen2016a} become gravitationally unstable but do not fragment. This result is likely attributed to their stricter sink particle creation algorithm while our production of many low-mass companions may be due to our more lenient algorithm. In this work, sink particles are created following the algorithm described in \citet{Krumholz2004a}, which allow sinks to form in regions that are Jeans unstable on the finest level and are undergoing gravitational collapse. The sink particle algorithm employed in \citet{Klassen2016a} includes these requirements but also enforces additional criteria following the work of \citet{Federrath2010b}. These additional requirements for sink particle creation are: (1) the flow must be convergent, (2) the location at which a sink can form must be a minimum for the gravitational potential, and (3) the total energy within a control volume around the potential sink particle be negative ($E_{\rm grav}+E_{\rm th}+E_{\rm kin}<0$). These additional requirements are essentially untested in the context of an unstable accretion disk, and it is not clear if they help prevent artificial fragmentation, or if they suppress the formation of fragments that would in fact form if the simulation had been carried out at higher resolution.

In contrast to our hierarchical systems, \citet{Krumholz2009a} form a massive binary with a mass ratio of $q = m_2/m_1 \approx 0.7$.  One key difference between the work of \citet{Krumholz2009a} and run \noturb\ presented in this work, besides the method of radiative transfer used, is the criterion used for merging sink particles. As described in Section \ref{sec:numerics} we allow two sink particles to merge when the lower mass particle has a mass less than 0.05 $M_{\rm \odot}$. \citet{Krumholz2009a} use the same sink particle creation algorithm as our work but allow particles to merge once two particles are within an accretion radius (i.e., 4 cells) of one another regardless of their mass. This lenient merging criteria allows for a series of disk-borne companions to merge together to form a more massive companion star, eventually resulting in a massive binary system. Furthermore, as run \nort\ demonstrated, the disk properties are highly dependent on the radiative transfer method employed. When the direct radiation pressure is neglected, we are left with hotter accretion disks that are less prone to fragmentation. The hotter gas yields higher accretion rates onto the companion stars, leading to larger masses.

Comparison of our results with previous work demonstrates that the multiplicity and companion mass distribution produced in massive star formation simulations is highly sensitive on the simulation resolution, radiative transfer, and algorithms used to create and merge sink particles. Since the fragmentation and resulting system multiplicity is sensitive to the numerics, we advise the reader that the fragmentation produced in our idealized numerical simulations and other simulations does not provide an adequate solution to the multiplicity of observed massive stars. Future work must include additional physics, such as magnetic fields and outflows, to further understand the observed multiplicity of massive stellar systems.

\subsection{Revisiting the Flashlight Effect}
\label{sec:flash}
In agreement with previous work, we find that disk accretion supplies most of the mass to the primary star in run \noturb. The presence of an optically thick accretion disk shields the stellar radiation field leading to the ``flashlight" effect in which the radiative flux escapes along the polar axis and into the polar cavities, launching radiation pressure dominated bubbles above and below the star \citep{Yorke2002a, Krumholz2009a, Kuiper2011a, Kuiper2012a, Klassen2016a}. As the bubbles expand they become unstable and deliver material to the star-disk system that can then be accreted onto the star. The asymmetry induced by the flashlight effect allows radiative flux to escape while mass from the disk can be accreted onto the star. 

In run \turb, however, we find that the flashlight effect is less important and that the initial turbulence allows for asymmetry in the radiation field. Instead of extended disk accretion the majority of the mass is delivered to the star by dense filaments and RT instabilities that are not destroyed by radiation pressure. The radiation freely escapes through low density channels that are not necessarily located along the polar directions of the primary star. The rapid infall of dense material inhibits the growth and stability of radiation pressure dominated bubbles at early times thus only allowing the ``flashlight" effect to occur at late times when the star has a larger luminosity. Therefore, we find that if the matter distribution of the core is asymmetric to begin with then the flashlight effect is not necessary for the formation of massive stars out of turbulent cores. However, our simulations neglect the effect of bipolar stellar outflows that have been shown to enhance the flashlight effect \citep{Cunningham2011a,Kuiper2015a}. 

For example, \citet{Cunningham2011a} were the first to perform 3D radiation hydrodynamic simulations of the collapse of initially turbulent pre-stellar cores that included feedback from bi-polar stellar outflows and radiation, but they neglected the direct radiation field. They found that inclusion of bipolar outflows from the massive protostar increases the ejection of ambient material along the polar directions of the star, enhancing the flashlight effect. This effect is enhanced because regions of the core that are expected to experience a large net force by the outward radiation force lie within the outflow cavity. The stellar outflows evacuate material along the polar directions of the primary star.  These outflow cavities provide significant focusing of the radiative flux in the polar directions, resulting in the radiative flux to escape while accretion continues through regions of the infalling envelope uninhibited by the radiative force.

\section{Conclusions}
\label{sec:conclusion}
In this paper, we have used our powerful new hybrid radiation transfer tool, \harm, in a suite of radiation hydrodynamic simulations that followed the collapse of initially laminar and turbulent massive pre-stellar cores to study the formation of massive stars. \harm\ uses a multi-frequency adaptive long-characteristics ray tracing scheme to capture the first absorption of the direct radiation from stars by the intervening interstellar dust and molecular gas, and flux limited diffusion to model the diffuse radiation field associated with the subsequent re-emission by interstellar dust. Our method is highly optimized and can run efficiently on hundreds of processors, works on adaptive grids, can be coupled to any moment method, and can be used for an arbitrary number of moving stars \citep{Rosen2016a}. 

The primary goal of our work is to determine how massive stars attain their mass when radiation pressure is the only feedback mechanism considered (i.e., in the absence of magnetic fields, outflows, and photoionization). Do massive stars obtain their mass through disk accretion alone? Or do radiative Rayleigh Taylor instabilities that develop in the radiation pressure dominated bubble shells, which are launched by the stars' intense radiation fields, deliver material directly via collapse onto the stars or star-disk systems? Or is it a combination of both of these processes?

For initially laminar cores, we find that the majority of mass delivered to the massive star is due to disk accretion, but that RT instabilities are responsible for bringing material onto the disk before it is subsequently incorporated into the star. For initially weakly turbulent cores, in contrast, we find that dense filaments and RT unstable material supply most of the mass to the accreting massive star directly, without mediation by an extended disk (i.e., an accretion disk with a radius larger than the 80 AU accretion zone radius of the sink particle) for the run time considered. However, we find that once an extended disk formed, disk accretion supplies material to the primary star. Our results show that the radiation escapes through low density channels that are not necessarily located along the polar directions of the star and that sustained radiation pressure dominated bubbles do not appear until late times when a significant accretion disk develops. For stronger turbulence at the level seen in many massive cores, we would expect this effect could be enhanced. Our results suggest that the ``flashlight" effect that occurs in our laminar runs, which allows the radiative flux to escape along the polar directions of the star and material to be accreted onto the star by an optically thick accretion disk, is not required for massive stars that form from turbulent cores. Instead, the asymmetric density distribution allows the radiation to escape through the path(s) of least resistance, allowing the dense infalling material to fall onto the star regardless of its location.

Our results also demonstrate that RT instabilities are a natural occurrence in the formation of massive stars regardless of whether the star-forming core is initially turbulent or laminar. These instabilities arise immediately for turbulent cores because the initial turbulence seeds the instabilities. RT instabilities develop later for laminar cores because the initially symmetric gas distribution must be perturbed. These perturbations can then seed RT instabilities that grow in time and can eventually deliver material to the star-disk system. We find that the development of an accretion disk and gravitational torques induced within the disk destroy the symmetry of the gas distribution and seed the initial perturbations that lead to RT instabilities in the bubble shells as first shown by \citet{Krumholz2009a}. Our work suggests that the seeds for RT instabilities that arise in initially laminar pre-stellar cores are asymmetries induced by the shielding of the direct radiation field by the accretion disk and the non-symmetric distribution of material within the bubbles. These asymmetries arise from disk flaring, disk fragmentation, and the gravitational interaction of the massive star with the accretion disk and companions. 

Previous work that simulated the collapse of initially laminar cores concluded that the direct radiation field inhibited the growth of RT instabilities \citep{Kuiper2012a, Klassen2016a}. Contrary to their results we find that inclusion of the direct radiation field only suppresses the non-linear growth of these instabilities at early times. As the asymmetry in the system grows, these instabilities can grow non-linearly and become dense enough to overcome the radiation-pressure barrier and deliver material to the star-disk system. We argue instead that poor shell resolution is the likely culprit as to why \citet{Kuiper2012a} and \citet{Klassen2016a} do not obtain bubble shells that become RT unstable. We check this hypothesis directly by conducting a resolution study where we intentionally de-resolve the bubble shell to the point where our resolution is comparable to that used in earlier work, and we show that doing so both delays the onset of instability and reduces its intensity. We further find in the work of \citet{Kuiper2012a}, that limitations of their fixed grid approach with a star that is centrally fixed in a spherical grid does not permit the movement of the star-disk system that would naturally allow asymmetries to arise and lead to seeding the RT instability.

We find that both turbulent and laminar cores lead to hierarchical star systems that consist of a massive primary star and several low-mass companions. We find that our multiplicity results are sensitive to the physics included, radiative transfer treatment used, and sink creation and merging criteria employed. Inclusion of the direct radiation pressure leads to cooler disks that are prone to greater fragmentation when compared to our comparison run that neglected the direct radiation field and assumed that the stellar radiation was immediately absorbed within the vicinity of the stars. However, given the idealized nature of our simulations, we cannot address the true multiplicity properties of massive stars.

Despite this limitation, our work settles a long-debated question in massive star formation: how does radiation pressure limit the masses of stars? We find that radiation pressure is still an important feedback mechanism that must be considered in massive star formation, but RT instabilities can overcome the radiation pressure barrier, at least in the context of the idealized numerical experiments performed thus far. However, there are still many other physical processes at play that we neglect. These include collimated outflows, fast stellar winds, and magnetic fields. Future work will include these other feedback mechanisms to determine a more complete picture of how massive stars form and how their associated feedback can limit stellar masses.

\section*{Acknowledgements}
\noindent
We thank Andrew Myers and Pak Shing Li for useful discussions regarding the \orion\ simulation code. ALR thanks Rolf Kuiper, Beno\^it Commer\c con, and Kaitlin Kratter for useful discussions about the results of our work. ALR and MRK acknowledge support from the National Aeronautics and Space Administration (NASA) through Hubble Archival Research grant HST-AR-13265.02-A issued by the Space Telescope Science Institute, which is operated by the Association of Universities for Research in Astronomy, Inc., under NASA contract NAS 5-26555 and Chandra Theory Grant Award Number TM5-16007X issued by the Chandra X-ray Observatory Center, which is operated by the Smithsonian Astrophysical Observatory for and on behalf of NASA under contract NAS8-03060. ALR acknowledges support from the NSF Graduate Research Fellowship Program and the AAUW American Fellowship Program. CFM and RIK acknowledge support from NASA through ATP grant NNX13AB84G and the NSF through grant AST-1211729.  RIK acknowledges the US Department of Energy at the Lawrence Livermore National Laboratory under contract DE-AC52-07NA27344. MRK, CFM, and RIK acknowledge support from NASA TCAN grant NNX-14AB52G. MRK acknowledges support from Australian Research Council grant DP160100695. This research was undertaken with the assistance of resources from the National Computational Infrastructure (NCI), which is supported by the Australian Government.

\bibliographystyle{mnras}
\bibliography{hmDR_alr_final.bbl}
\end{document}